\documentclass[journal]{IEEEtran}
\usepackage{amsmath,amsfonts,amssymb}
\usepackage{algorithmic}
\usepackage{algorithm}
\usepackage{color,array,amsthm}
\usepackage[caption=false,font=normalsize,labelfont=sf,textfont=sf]{subfig}
\usepackage{textcomp}
\usepackage{stfloats}
\usepackage{url}
\usepackage{verbatim}
\usepackage{graphicx}
\usepackage{cite}
\usepackage{siunitx}
\usepackage{microtype}

\newtheorem{lemma}{Lemma}
\newtheorem{remark}{Remark}
\newtheorem{definition}{Definition}
\newtheorem{proposition}{Proposition}

\allowdisplaybreaks

\begin{document}

\title{Probabilistic GOSPA: A Metric for Performance Evaluation of Multi-Object Filters with Uncertainties}

\author{Yuxuan Xia,~\IEEEmembership{Member,~IEEE,} 
{\'A}ngel F. Garc{\'i}a-Fern{\'a}ndez,
Johan Karlsson~\IEEEmembership{Senior Member,~IEEE,}\\  
Kuo-Chu Chang,~\IEEEmembership{Life Fellow,~IEEE,} 
Ting Yuan,~\IEEEmembership{Senior Member,~IEEE,}
Lennart Svensson,~\IEEEmembership{Senior Member,~IEEE}
\thanks{Yuxuan Xia and Ting Yuan are with the Department of Automation and Intelligent Sensing, Shanghai Jiao Tong University, Shanghai, China  (e-mail: yuxuan.xia@sjtu.edu.cn, tyuan@sjtu.edu.cn).}
\thanks{{\'A}ngel F. Garc{\'i}a-Fern{\'a}ndez is with the IPTC, ETSI de Telecomunicación (ETSIT), Universidad Politécnica de Madrid, Madrid, Spain (e-mail: angel.garcia.fernandez@upm.es).}
\thanks{Johan Karlsson is with the Department of Mathematics, KTH Royal Institute of Technology, Stockholm, Sweden (e-mail: johan.karlsson@math.kth.se).}
\thanks{Kuo-Chu Chang is with SEOR, George Mason University, Fairfax, VA, USA (e-mail: kchang@gmu.edu).}
\thanks{Lennart Svensson is with the Department of Electrical Engineering, Chalmers University of Technology, Gothenburg, Sweden (e-mail: lennart.svensson@chalmers.se).}
\thanks{This work was partially funded by SJTU-KTH collaboration and research development seed grants.}}



\maketitle

\begin{abstract}
    This paper presents a probabilistic generalization of the Generalized Optimal Sub-Pattern Assignment (GOSPA) metric, termed P-GOSPA. The GOSPA metric has been widely used to evaluate the distance between finite sets, particularly in multi-object estimation applications. The P-GOSPA extends GOSPA into the space of multi-Bernoulli densities, incorporating inherent uncertainty in probabilistic multi-object representations. Additionally, P-GOSPA retains the interpretability of GOSPA, such as its decomposition into localization, missed detection, and false detection errors in a sound and meaningful manner. Examples and simulations are provided to demonstrate the efficacy of the proposed P-GOSPA metric.
\end{abstract}


\section{Introduction}

Multi-object tracking (MOT) involves sequentially estimating the states of moving objects, which may enter or leave the surveillance area, given noisy sensor measurements \cite{blackman1999design}. When developing and testing different MOT algorithms in varying scenarios, it is crucial to assess and compare their performances. To achieve this, a reliable performance metric is needed to measure the distance between the ground truth and the estimates.

Early MOT evaluation methods rely on intuitive concepts, including localization errors for properly detected objects, as well as missed and false object detection errors \cite[Sec. 13.6]{blackman1999design}, \cite{fridling1991performance,drummond1992ambiguities,mabbs1993performance,rothrock2000performance}. However, these methods typically rely on ad hoc mechanisms. Later, mathematically sound MOT performance evaluation methods have been developed based on finite sets~\cite{mahler2014advances}, including the optimal mass transfer (OMAT) metric~\cite{hoffman2004multitarget} and the Hausdorff metric \cite{hoffman2004multitarget}. These methods measure the distance between the sets of ground truth objects and estimated object states according to a mathematically well-defined metric. However, as discussed in \cite{schuhmacher2008consistent}, these two metrics both entail a host of drawbacks. Most notably, the Hausdorff metric is insensitive to cardinality mismatches, whereas OMAT lacks a physically consistent interpretation when the cardinalities of multi-object estimates differ.

To address the limitations of the OMAT and Hausdorff metrics, the Optimal Sub-Pattern Assignment (OSPA) metric was presented in \cite{schuhmacher2008new,schuhmacher2008consistent}. The OSPA metric calculates normalized localization errors for optimally associated ground truth and estimated object states while also incorporating penalties for cardinality mismatches. Later, an extension of OSPA without normalization was used in \cite{barrios2016metrics}. While OSPA is mathematically sound and has more reasonable interpretations in terms of varying cardinalities, unlike traditional MOT evaluation methods, it does not penalize missed and false detection errors based on intuitive concepts \cite[Sec. 13.6]{blackman1999design}, \cite{fridling1991performance,drummond1992ambiguities,mabbs1993performance,rothrock2000performance}. 

A metric, that can quantify all the above aspects in a mathematically consistent way, is the generalized OSPA (GOSPA) metric, proposed in \cite{rahmathullah2017generalized}. 
Importantly, GOSPA penalizes localization errors for properly detected objects as well as missed and false detection errors. Additionally, GOSPA eliminates the spooky effect observed in optimal multi-object estimation with OSPA \cite{garcia2019spooky}, and it has also shown advantages over OSPA in sensor management \cite{garcia2021analysis}.

Most MOT algorithms rely on recursive Bayesian estimation, where, at each time step, estimated object states are extracted from the multi-object posterior density. The MOT performance is then evaluated by computing the distance between the sets of ground truth objects and estimated object states using metrics such as OSPA or GOSPA. Clearly, this performance evaluating procedure does not account for the uncertainty information in multi-object posterior densities. One way to achieve this is to compute the mean (G)OSPA distance, averaged over the posterior density. However, computing the mean (G)OSPA is often a non-trivial task, due to the lack of analytical expressions.

In the literature, there have been only a few attempts trying to (partially) assess the multi-object filtering performance with uncertainties in an efficient and tractable way. In \cite{nagappa2011incorporating}, object state estimation uncertainties (covariances) are integrated into OSPA by using the Hellinger distance as the base distance, which has a closed-form expression for Gaussian distributions. In \cite{he2013track}, a quality-based OSPA (Q-OSPA) was presented by incorporating object existence uncertainties into OSPA. However, as we will demonstrate later in Section \ref{sec_pgospa}.\ref{subsec_pgospa_interpretation}, Q-OSPA is not a mathematically well-defined metric because it fails to satisfy the definiteness property. Furthermore, it often lacks a reasonable physical interpretation, making it less suitable for practical applications. More recently, the negative log-likelihood of the multi-object posterior given the ground truth object states, was proposed in \cite{pinto2021uncertainty} as a performance measure. However, it is not a true metric and can yield infinite values for certain multi-object densities\footnote{For example, when evaluating the set of ground truth objects at a multi-
Bernoulli density whose number of Bernoulli components is smaller than the
number of ground truth objects.}, making it difficult to consistently distinguish between estimation results.

\begin{figure}[!t]
\centerline{\includegraphics[width=\linewidth]{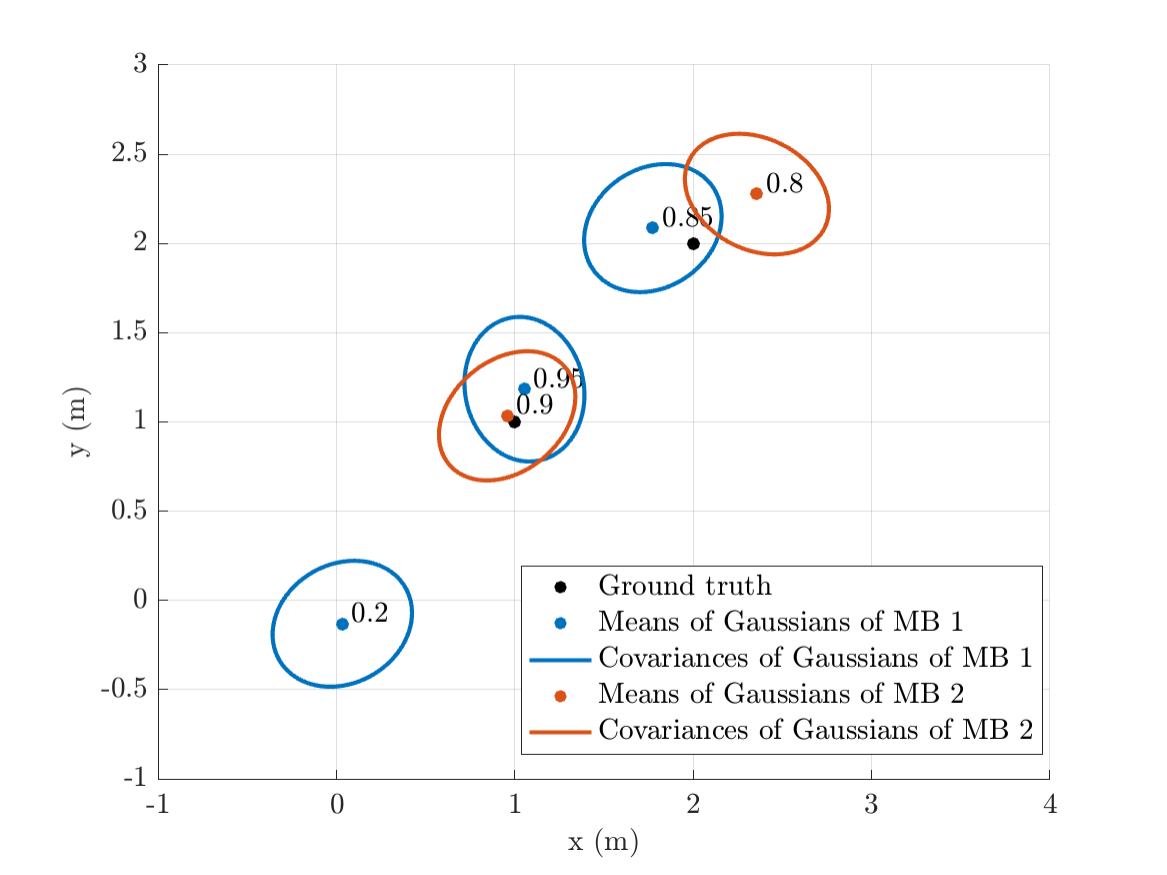}}
\caption{An exemplary scenario with two objects and two MB set densities. Each Bernoulli density has Gaussian single-object density, and its existence probability is shown next to its Gaussian mean. A desirable metric should be able to answer: 1) what is the distance between each MB density and ground truth object states? and 2) what is the distance between the two MB densities?}
\label{fig_ex1}
\end{figure}

Many widely used MOT algorithms, such as those based on multi-Bernoulli (MB) conjugate priors \cite{garcia2018poisson,williams2015marginal,garcia2019mbm,reuter2014labeled}, can generate sets of state estimates representing potential objects. These estimates account for both existence uncertainty and state estimation uncertainty, encapsulated in the form of an MB density. In addition, the set of ground truth object states can also be regarded as an MB density, where all the Bernoulli components have probability of existence one and Dirac delta single-object densities. This also holds in simulations, where sets of ground truth objects are often obtained by sampling from a multi-object set density. For a standard multi-object dynamic model with MB birth \cite{mahler2014advances}, the multi-object density takes the form of an MB distribution. Therefore, it is desirable to have a metric that can fully account for the uncertainty information captured by an MB set density, as illustrated in Fig. \ref{fig_ex1}. 
Furthermore, the metric should be mathematically well-defined, physically
interpretable, and computationally practical.

In this paper, we present such a metric, which can be considered as a probabilistic generalization of the GOSPA (P-GOSPA) metric. While GOSPA directly measures distances between deterministic sets, P-GOSPA operates on distributions over MB random finite sets, enabling a principled evaluation of performance in probabilistic contexts. We also show that P-GOSPA inherits the interpretability of GOSPA. For a specific choice of its parameters, P-GOSPA can be decomposed into four components: the expected localization error and existence probability mismatch error for correctly detected objects, along with the expected missed and false detection errors. In addition to MOT performance evaluation, we also demonstrate in the simulations that P-GOSPA can be used to quantify the approximation errors in recursive MOT filtering.

The rest of the paper is structured as follows. The P-GOSPA metric is presented in Section II. Illustrative examples and simulation results are shown in Section III and Section IV, respectively. Conclusions are summarized in Section V.

\section{Probabilistic GOSPA Metric}\label{sec_pgospa}

In this section, we first introduce the MB set density, and then we present the P-GOSPA metric, which measures the distance between two MB set densities.

A Bernoulli process $X$ is a random finite set with a Bernoulli-distributed cardinality $|X|$, and its density is 
\begin{equation}
    f(X) = \begin{cases}
        1 - r & X = \emptyset \\
        rp(x) & X = \{x\} \\
        0 & \text{otherwise},
    \end{cases}
\end{equation}
where $r\in[0,1]$ is the probability of existence, and $p(\cdot)$ is the single-object density conditioned on object existence. The single object state $x \in \mathbb{X}$ belongs to the object space $\mathbb{X}$, which is locally compact, Hausdorff and second-countable \cite{mahler2014advances}. An MB process with $n$ Bernoulli components is a disjoint union of $n$ independent Bernoulli processes. Assume that the $i$-th Bernoulli component is parameterized by $r^i$ and $p^i(\cdot)$, where $i\in\{1,\dots,n\}$, then the density of the MB process, consisting of these Bernoulli components, can be completely described by parameters $\{(r^i,p^i(\cdot))\}_{i=1}^n$.

To introduce P-GOSPA, we first define the two MB densities on which it is evaluated.
Let $f_X(\cdot)$ and $f_Y(\cdot)$ be two MB densities: $f_X(\cdot)$ has $n_X$ Bernoulli components, where the $i$-th Bernoulli component is parameterized by existence probability $r_x^i \in (0,1]$ and single-object density $p_x^i(\cdot)$; and $f_Y(\cdot)$ has $n_Y$ Bernoulli components, where the $j$-th Bernoulli component is parameterized by existence probability $r_y^j \in (0,1]$ and single-object density $p_y^j(\cdot)$. 

\begin{definition}\label{defi_pgospa}
    Let $d(p_x,p_y)$ denote a metric for single-object densities $p_x(x)$ and $p_y(y)$ for any single-object states $x,y\in\mathbb{X}$, and let $d^{(c)}(p_x,p_y) = \min(d(p_x,p_y),c)$, where $c > 0$, denote the cut-off metric of $d(p_x,p_y)$. Let $\Pi_n$ denote the set of all permutations of $\{1,\dots,n\}$ for any $n\in\mathbb{N}$, where each element $\pi\in \Pi_n$ be a sequence $(\pi(1),\dots,\pi(n))$. For $n_X \leq n_Y$, the P-GOSPA metric is defined as\footnote{The expression of P-GOSPA is also valid for MB set densities having Bernoulli components with zero existence probability. In fact, it is easy to verify that P-GOSPA remains unchanged if an arbitrary number of Bernoulli components with zero existence probability are appended to one or both MB set densities. This makes sense as Bernoulli components with zero existence probability carry no uncertainty information. Nevertheless, we restrict the existence probability of Bernoulli components to $(0,1]$ to ensure that P-GOSPA satisfies the definiteness property.}
    \begin{align}
        &d_p^{(c,\alpha)}(f_X,f_Y) \nonumber \\ &\triangleq \left[\min_{\pi \in \Pi_{n_Y}}\left(\sum_{i=1}^{n_X}\left[\min\left(r_x^i,r_y^{\pi(i)}\right)d^{(c)}\left(p^i_x,p_y^{\pi(i)}\right)^p \right.\right.\right. \nonumber \\
        &~~~\left.\left.\left. + \left|r_x^i - r_y^{\pi(i)}\right|\frac{c^p}{\alpha}\right] + \frac{c^p}{\alpha}\sum_{i = n_X + 1}^{n_Y}r_y^{\pi(i)}\right)\right]^{\frac{1}{p}}, \label{eq_pgospa}
    \end{align}
    where $0 < \alpha \leq 2$ and $1 \leq p < +\infty$. If $n_X > n_Y$, $d_p^{(c,\alpha)}(f_X,f_Y) \triangleq d_p^{(c,\alpha)}(f_Y,f_X)$.
\end{definition}
It follows directly from the definition that P-GOSPA satisfies the non-negativity, definiteness and the symmetry properties of a metric. The proof of the triangle inequality is provided in Appendix \ref{appendix_triangle_pgospa}. The roles of the parameters $c$, $p$, and $\alpha$ in P-GOSPA are similar to those in GOSPA; they will be elaborated later in Section \ref{sec_pgospa}.\ref{subsec_pgospa_interpretation}.

We note that if all the Bernoulli components in $f_X(\cdot)$ and $f_Y(\cdot)$ have existence probability one and Dirac delta single-object densities, P-GOSPA reduces to the original GOSPA metric between the finite sets $X= \{x_1,\dots,x_{n_X}\}$ and $Y= \{y_1,\dots,y_{n_Y}\}$ \cite{rahmathullah2017generalized},
\begin{align}
    &\bar{d}_p^{(c,\alpha)}(X,Y) \nonumber \\ &\triangleq \left[\min_{\pi \in \Pi_{n_Y}}\sum_{i=1}^{n_X}\bar{d}^{(c)}\left(x_i,y_{\pi(i)}\right)^p  + \frac{c^p}{\alpha}(n_Y-n_X)\right]^{\frac{1}{p}}, \label{eq_gospa}
\end{align}
if $n_X \leq n_Y$, and $\bar{d}_p^{(c,\alpha)}(X,Y) \triangleq \bar{d}_p^{(c,\alpha)}(Y,X)$ if $n_X > n_Y$, where $\bar{d}^{(c)}(x,y)$ is the cut-off distance of $\bar{d}(x,y) = d(\delta_x(\cdot),\delta_y(\cdot))$ for Dirac delta densities $\delta_x(\cdot)$, $\delta_y(\cdot)$, centred at $x,y\in\mathbb{X}$, respectively. Therefore, P-GOSPA can be considered as a probabilistic generalization of GOSPA, incorporating the uncertainties in the MB set densities.

\begin{remark}
GOSPA has been extended in \cite{rahmathullah2017generalized} to ground truth and object state estimates that are random finite sets, and it is called the average GOSPA metric \cite[Prop. 2]{rahmathullah2017generalized}, defined as $\mathbb{E}[\bar{d}_p^{(c,\alpha)}(X,Y)^{p^\prime}]^{1/p^\prime}$ where $p^\prime < \infty$. We note that the average GOSPA metric does not have an analytical expression in general, even if we assume that both the sets of ground truth objects and estimated objects have independent MB set densities. As a comparison, P-GOSPA has an analytical expression \eqref{eq_pgospa} and can be efficiently computed by solving a 2D assignment problem using, for example, the JVC algorithm~\cite{crouse2016implementing}.

\end{remark}

\subsection{Interpretation of P-GOSPA}\label{subsec_pgospa_interpretation}

We briefly discuss the roles of $d(p_x,p_y)$ and parameters $p$, $c$ and $\alpha$ used in P-GOSPA. Specifically, $d(p_x,p_y)$ is a metric between two probability densities on the space $\mathbb{X}$, including, e.g., Wasserstein distance, and Hellinger distance. As a comparison, the distance metric $\bar{d}(x,y)$ in GOSPA is defined on single object space $\mathbb{X}$. The selection of $d(p_x,p_y)$ in P-GOSPA and its connection to $\bar{d}(x,y)$ in GOSPA will be discussed later in Section \ref{sec_example}.

The maximum allowable distance between two single-object densities is given by the cut-off distance $c$. The exponent $p$ plays a similar role as in GOSPA: larger values of $p$ impose stronger penalties on outliers. The parameters $p$, $c$ and $\alpha$ collectively determine the penalization of the expected cardinality mismatch, with lower values of $\alpha$ increasing the cost. The expected cardinality mismatch for $n_X \leq n_Y$ is
\begin{equation*}
    \sum_{i=1}^{n_X}\left|r^i_x - r^{\pi(i)}_y\right| + \sum_{i=n_X+1}^{n_Y}r_y^{\pi(i)}.
\end{equation*}

From \eqref{eq_pgospa}, we observe that P-GOSPA consists of two terms: one accounting for
the costs of associated pairs of Bernoulli components, and the other capturing
the costs of unassigned components in the MB density with a larger number of
Bernoulli components.
\begin{lemma}\label{lemma_pgospa_bernoulli}
    The P-GOSPA metric between two Bernoulli set densities $f_X(\cdot)$ and $f_Y(\cdot)$, parameterized by existence probability $r_x$ and $r_y$, and single-object densities $p_x(\cdot)$ and $p_y(\cdot)$, respectively, is 
    \begin{align}
        &d_p^{(c,\alpha)}(f_X,f_Y) \nonumber\\
        &\triangleq \left(\min(r_x,r_y)d^{(c)}(p_x,p_y)^p + |r_x - r_y|\frac{c^p}{\alpha}\right)^{\frac{1}{p}}. \label{eq_bernoulli_cost}
    \end{align}
\end{lemma}
Lemma \ref{lemma_pgospa_bernoulli} is a special case of the P-GOSPA metric \eqref{eq_pgospa} with $n_X = n_Y = 1$. It can be seen from \eqref{eq_bernoulli_cost} that the error between two Bernoulli densities can be decomposed into the expected localization error and the existence probability mismatch error, represented by the first and the second term in \eqref{eq_bernoulli_cost}, respectively. While it is reasonable to see that the existence probability mismatch error depends on $|r_x - r_y|$, it is not straightforward to build intuition on why the expected localization error is influenced by $\min(r_x,r_y)$. A reasonable explanation is provided in the following Lemma.

\begin{lemma}\label{lemma_wasserstein_bernoulli}
    Let the two Bernoulli set densities $f_X(\cdot)$ and $f_Y(\cdot)$ have Dirac delta single-object densities, and $d(\delta_x(\cdot),\delta_y(\cdot)) = \bar{d}(x,y)$ be the $p$-Wasserstein distance between point masses $x$ and $y$. Then the P-GOSPA metric between the two Bernoulli set densities is the same as the $p$-Wasserstein distance $W_p(f_X,f_Y)$, using GOSPA as its cost function between them, i.e.,
    \begin{align}
        &W^p_p(f_X,f_Y)\nonumber \\
        &\triangleq \inf_{q\in\mathcal{Q}(f_X,f_Y)}\iint \bar{d}^{(c,\alpha)}_p(X,Y)^p q(X,Y) \delta X \delta Y \nonumber\\ 
        &= \min(r_x,r_y)\bar{d}^{(c)}(x,y)^p + |r_x - r_y|\frac{c^p}{\alpha}, \label{eq_bernoulli_wasserstein}
    \end{align}
    where $\mathcal{Q}(f_X,f_Y)$ denotes the set of all the joint distributions $q$ for $(X,Y)$ that have marginals $f_X(\cdot)$ and $f_Y(\cdot)$, respectively, and $\int f(\cdot)\delta X$ denotes the set integral \cite{mahler2014advances}.
\end{lemma}


We note that $f_X(\cdot)$ and $f_Y(\cdot)$ only provide marginal distributions of the Bernoulli sets $X$ and $Y$. The Wasserstein distance computes the expected GOSPA under the joint distribution $q(X,Y)$ that yields the smallest value. Naturally, Bernoulli sets $X$ and $Y$ should ideally either both be empty or
both be non-empty, especially when $x$ and $y$ are close, giving rise to the term $\min(r_x,r_y)$.

Lemma \ref{lemma_wasserstein_bernoulli} is based on the assumption that the single-object densities $p_x(\cdot)$ and $p_y(\cdot)$ are Dirac delta functions. If this assumption is removed, P-GOSPA can then be interpreted as an upper bound on the $p$-Wasserstein distance $W_p(f_X,f_Y)$. This relationship is formalized in the following Proposition.

\begin{proposition}\label{prop_Wasserstein_bernoulli}
    Let $d(p_x,p_y)$ be the $p$-Wasserstein distance between single-object densities $p_x(\cdot)$ and $p_y(\cdot)$. The P-GOSPA metric between two Bernoulli set densities can be interpreted as an upper bound on the $p$-Wasserstein distance, using GOSPA as its cost function between them. 
\end{proposition}


Proposition \ref{prop_Wasserstein_bernoulli} is proved in Appendix \ref{appendix_proposition_Wasserstein_bernoulli}\footnote{The upper bound in Proposition \ref{prop_Wasserstein_bernoulli} comes from the fact that the integral of the minimum is always less than or equal to the minimum of the integrals; the detailed derivation can be found in Appendix \ref{appendix_proposition_Wasserstein_bernoulli}.}. Lemma \ref{lemma_wasserstein_bernoulli} is a special case of Proposition \ref{prop_Wasserstein_bernoulli} where the upper bound is lifted, and it is proved in Appendix \ref{appendix_lemma_wasserstein_bernoulli}. 

\begin{remark}
    The P-GOSPA metric can also be interpreted as a GOSPA metric applied to sets of Bernoulli densities. To do so, we first append Bernoulli densities with zero existence probability to the MB with lower number of Bernoulli components (which does not affect the P-GOSPA metric value). The base metric for GOSPA is then defined as in \eqref{eq_bernoulli_cost}, and the cut-off value can be set to any value greater than or equal to the maximum value of \eqref{eq_bernoulli_cost}, which is $c/\alpha^{1/p}$ for $0 < \alpha \leq 1$ and $c$ for $1 \leq \alpha \leq 2$.
\end{remark}

\begin{remark}
    The incorporation of object existence probabilities (also called track qualities) into MOT performance metric was considered in \cite{he2013track}, where the metric is called Q-OSPA. Q-OSPA extends OSPA by proposing a new base metric between deterministic object states with existence uncertainties. Specifically, for two single-object states $x$ and $y$ with existence probabilities $r_x$ and $r_y$, respectively, their distance is 
    \begin{equation}
        \tilde{d}^{(c)}(x,y) = r_xr_y\bar{d}^{(c)}(x,y) + (1-r_xr_y)c.
    \end{equation}
    This new base metric is not mathematically well-defined since it does not satisfy the definiteness property\footnote{For example, when $x = y$ (such that $\bar{d}^{(c)}(x,y) = 0$) and $r_x=r_y$, for any $r_x,r_y\in(0,1)$, $\tilde{d}^{(c)}(x,y) = (1-r_xr_y)c \neq 0$.}. Moreover, the term $(1-r_xr_y)c$ does not have intuitive physical interpretation. This is because $1-r_xr_y$ represents the probability that at least one of the object states does not exist, including the case where neither exists, which should not be penalized.
\end{remark}

\subsection{Motivation for setting $\alpha = 2$ in P-GOSPA for MOT}

Similar to GOSPA, setting $\alpha = 2$ in P-GOSPA is the most suitable choice for evaluating MOT algorithms. We demonstrate that, with this choice, the distance metric can be decomposed into expected association errors for properly detected objects (which have Bernoulli set densities), and expected missed and false detection errors, represented by Bernoulli components left unassigned.

Since P-GOSPA is symmetric, without loss of generality, we assume that $f_X(\cdot)$ and $f_Y(\cdot)$ are the ground truth and the estimated multi-object densities, respectively. We consider two Bernoulli components, one from $f_X(\cdot)$ and the other from $f_Y(\cdot)$, each with existence probabilities $r_x$, $r_y$, and single-object densities $p_x(\cdot)$, $p_y(\cdot)$, respectively. We further assume that $p_x(\cdot)$ and $p_y(\cdot)$ are sufficiently different from any of the single-object densities in $f_Y(\cdot)$ and $f_X(\cdot)$, respectively. Under this case, the ground truth Bernoulli density with $p_x(\cdot)$ is missed detected, and the Bernoulli component with $p_y(\cdot)$ represents the set density estimation of a false detected object. If none of the two Bernoulli components have been associated to any other Bernoulli components in the permutation $\pi$ in \eqref{eq_pgospa}, they, together, contribute with a cost $(r_x+r_y)c^p/\alpha$. If the two Bernoulli components are associated to each other in the permutation in \eqref{eq_pgospa} instead, the cost of contribution of this assignment is $\min(r_x,r_y)c^p + |r_x-r_y|c^p/\alpha$.

We argue that, for MOT performance evaluation, the cost of one missed detection and one false detection should, in principle, be equivalent to the cost of associating these two Bernoulli components with each other\footnote{In practice, the two costs may differ depending on the specific application or weighting scheme used, which could influence the interpretation of the results.}. That is, it should hold that 
\begin{equation}
    \frac{(r_x+r_y)c^p}{\alpha} = \min(r_x,r_y)c^p + \frac{|r_x-r_y|c^p}{\alpha},
\end{equation}
and in this case $\alpha = 2$. Therefore, $\alpha = 2$ in P-GOSPA is the most appropriate choice. From this point forward, we refer the term P-GOSPA to P-GOSPA with $\alpha = 2$.

In P-GOSPA, an unassigned Bernoulli component (either missed or false) with existence probability $r$ always costs $rc^p/2$. This suggests an alternative form of the P-GOSPA metric, consisting of expected association errors for properly detected objects (Bernoulli components) and costs for Bernoulli components left unassigned. Similar to GOSPA, this alternative expression of P-GOSPA can be reformulated in terms of 2D assignment functions. 

Specifically, let $\gamma \in \Gamma$ be an assignment set between $\{1,\dots,n_X\}$ and $\{1,\dots,n_Y\}$ with the following properties: $\gamma \subseteq \{1,\dots,n_X\}\times\{1,\dots,n_Y\}$, $(i,j),(i,j^\prime)\in\gamma \Rightarrow j = j^\prime$ and $(i,j),(i^\prime,j)\in\gamma \Rightarrow i = i^\prime$, where $\Gamma$ represents the set of all possible assignment sets. Using this, we can formulate the following proposition.

\begin{proposition}\label{prop_pgospa_alpha2}
    For $\alpha = 2$, the P-GOSPA metric can be formulated as an optimization problem over assignment sets,
    \begin{align}
        &d_p^{(c,2)}(f_X,f_Y)\nonumber\\ 
        &= \left[\min_{\gamma\in\Gamma}\left(\sum_{(i,j)\in\gamma}\left[\min\left(r_x^i,r_y^{j}\right)d\left(p_x^i,p_y^{j}\right)^p + \left|r_x^i-r_y^j\right|\frac{c^p}{2} \right] \right.\right.\nonumber \\
        &~~~~~\left.\left.+\frac{c^p}{2}\left(\sum_{i:\forall j,(i,j)\notin\gamma}r_x^i + \sum_{j:\forall i,(i,j)\notin\gamma}r_y^j\right) \right)\right]^{\frac{1}{p}}. \label{eq_assignment_pgospa}
    \end{align}
\end{proposition}
Proposition \ref{prop_pgospa_alpha2} is proved in Appendix \ref{appendix_proposition_pgospa_alpha2}.

Proposition \ref{prop_pgospa_alpha2} confirms that P-GOSPA penalizes unassigned objects and association errors for properly detected objects. Specifically, the $p$-th order P-GOSPA consists of four different terms:
\begin{itemize}
    \item $\sum_{(i,j)\in\gamma}\min(r_x^i,r_y^j)d(p_x^i,p_y^j)^p$: the expected localization error for associated Bernoulli components.
    \item $\sum_{(i,j)\in\gamma}|r_x^i-r_y^j|c^p/2$: the existence probability mismatch error for associated Bernoulli components.
    \item $c^p/2\sum_{i:\forall j,(i,j)\notin\gamma}r_x^i$: the expected missed object detection error.
    \item $c^p/2\sum_{j:\forall i,(i,j)\notin\gamma}r_y^j$: the expected false object detection error.
\end{itemize}
To understand this decomposition, we note that $|\gamma|$ is the number of pairs of associated Bernoulli components, and $i$, $j$ that are left unassigned represent indices of Bernoulli components representing missed and false detections. We also note that the concept of the cut-off distance $d^{(c)}(\cdot,\cdot)$ becomes irrelevant in \eqref{eq_assignment_pgospa}, as Bernoulli components with single-object densities that are significantly distant from the single-object densities of any other Bernoulli components remain unassigned.

\section{Illustrative Examples}\label{sec_example}

In this section, we give two illustrative examples to show how P-GOSPA can quantify the uncertainties represented by MB set densities. In the examples, we set $c = 5$, $p = 1$, and we use the $2$-Wasserstein distance, whose cost function is given by the Euclidean distance, as the base metric $d(p_x,p_y)$ between two Gaussian distributions $p_x(x) = \mathcal{N}(x;m_x,P_x)$ and $p_y(y) = \mathcal{N}(y;m_y,P_y)$ \cite{olkin1982distance},
\begin{align}
    &W_2(p_x,p_y)= \nonumber\\
    &\left[\Vert m_x-m_y\Vert^2_2 + \text{trace}\left(P_x + P_y - 2\left(P_y^{\frac{1}{2}}P_xP_y^{\frac{1}{2}}\right)^{\frac{1}{2}}\right)\right]^{\frac{1}{2}},
\end{align}
where $P_y^{1/2}$ is the principle square root of covariance $P_y$. If $y$ becomes a point mass, i.e., $p_y(y)$ is a Dirac delta function centered at $m_y$, the Wasserstein distance can be obtained by setting $P_y$ to zero, which yields
\begin{equation}
    W_2(p_x,p_y) = \left[\Vert m_x-m_y\Vert^2_2 + \text{trace}(P_x)\right]^{\frac{1}{2}}.
\end{equation}
This encodes both the mean discrepancy and the spread (uncertainty) of the Gaussian distribution $p_x(x)$. Moreover, when both $p_x(x)$ and $p_y(y)$ are Dirac delta functions, the Wasserstein distance between $p_x(x)$ and $p_y(y)$ reduces to the Euclidean distance between the point masses $x$ and $y$, which is the typical distance metric used in GOSPA.

Compared to the Hellinger distance used in \cite{nagappa2011incorporating} that is between 0 and 1, the Wasserstein distance has a more intuitive physical interpretation, and it also has an analytical expression for particle-based state representations. In addition, the Wasserstein distance has often been used to measure the state estimation errors for tracking extended objects with elliptical shapes \cite{yang2016metrics}.

\subsection{Example 1}

Let us consider a 1D example, where the true object state is at \SI{0}{\meter}, and the MB density has a single Bernoulli with existence probability $r$ and Gaussian density with mean \SI{2}{\meter} and variance $\sigma^2$. We further assume that the object state estimate \SI{2}{\meter} will be reported by the estimator only if $r \geq 0.5$, and that the base metric in GOSPA is the Euclidean distance. In this case, the GOSPA error is $\bar{d}_1^{(5,2)} = 2$ if $r \geq 0.5$ and $\bar{d}_1^{(5,2)} = 2.5$ if $r < 0.5$, whereas according to \eqref{eq_bernoulli_cost}, the P-GOSPA error is 
\begin{equation}
    d_1^{(5,2)} = \min(5,\sqrt{4+\sigma^2})r + 2.5(1-r). \label{eq_heatmap}
\end{equation}

\begin{figure}[!t]
    \centerline{\includegraphics[width=\linewidth]{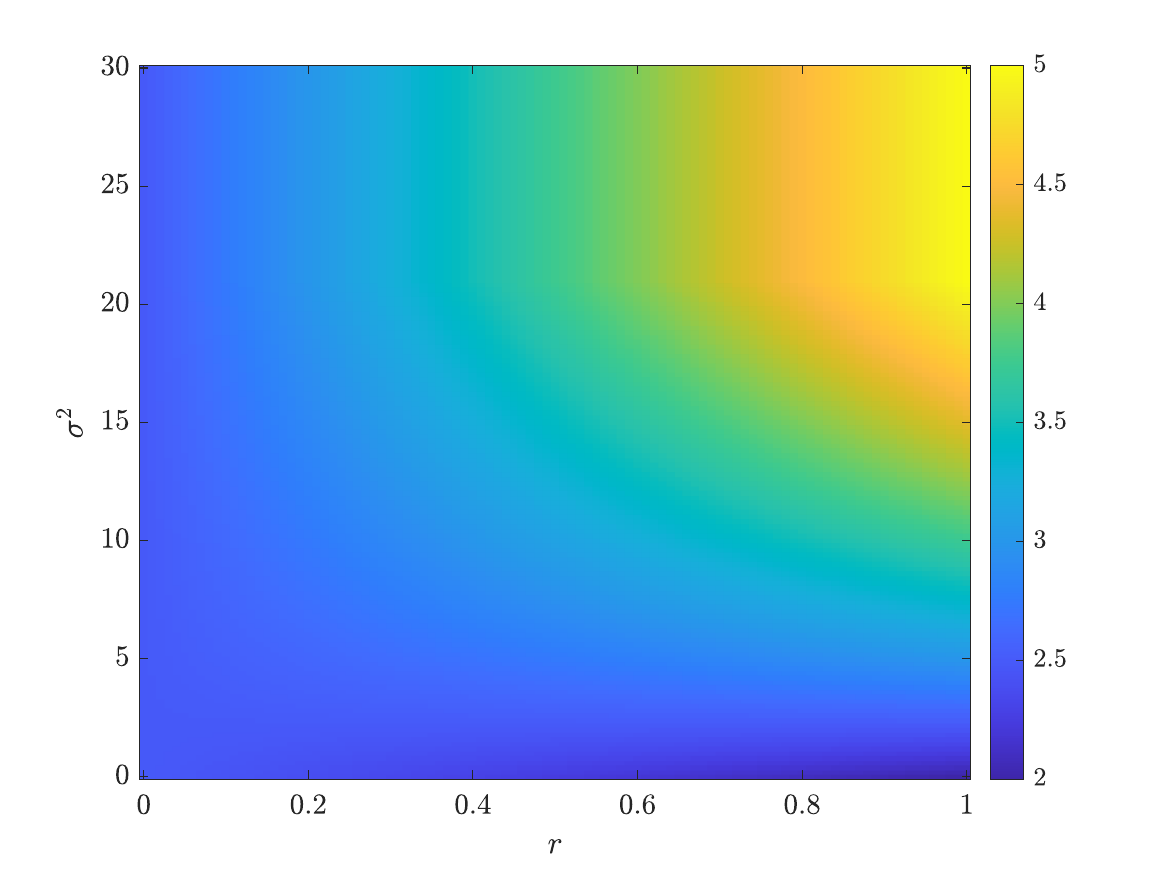}}
    \caption{Example 1: P-GOSPA versus $r$ and $\sigma^2$.}
    \label{fig_ex2}
\end{figure}

The heatmap representation of the P-GOSPA errors versus $r$ and $\sigma^2$, computed using \eqref{eq_heatmap}, is shown in Fig. \ref{fig_ex2}. We observe that P-GOSPA effectively accounts for the uncertainties in the Bernoulli density, including variations in existence probabilities and Gaussian variances, while maintaining an (almost) smooth transition in response to these changes. For the special case $r = 0$, P-GOSPA $d_1^{(5,2)}$ is a constant since for a non-existent object, its single-object density has no effect. We also note that when Gaussian variance $\sigma^2 \geq 21$, the true object becomes missed detected and its estimate becomes a false detection, and thus P-GOSPA $d_1^{(5,2)}$ becomes invariant to $\sigma^2$.

\subsection{Example 2}

\begin{figure}[!t]
    \centerline{\includegraphics[width=\linewidth]{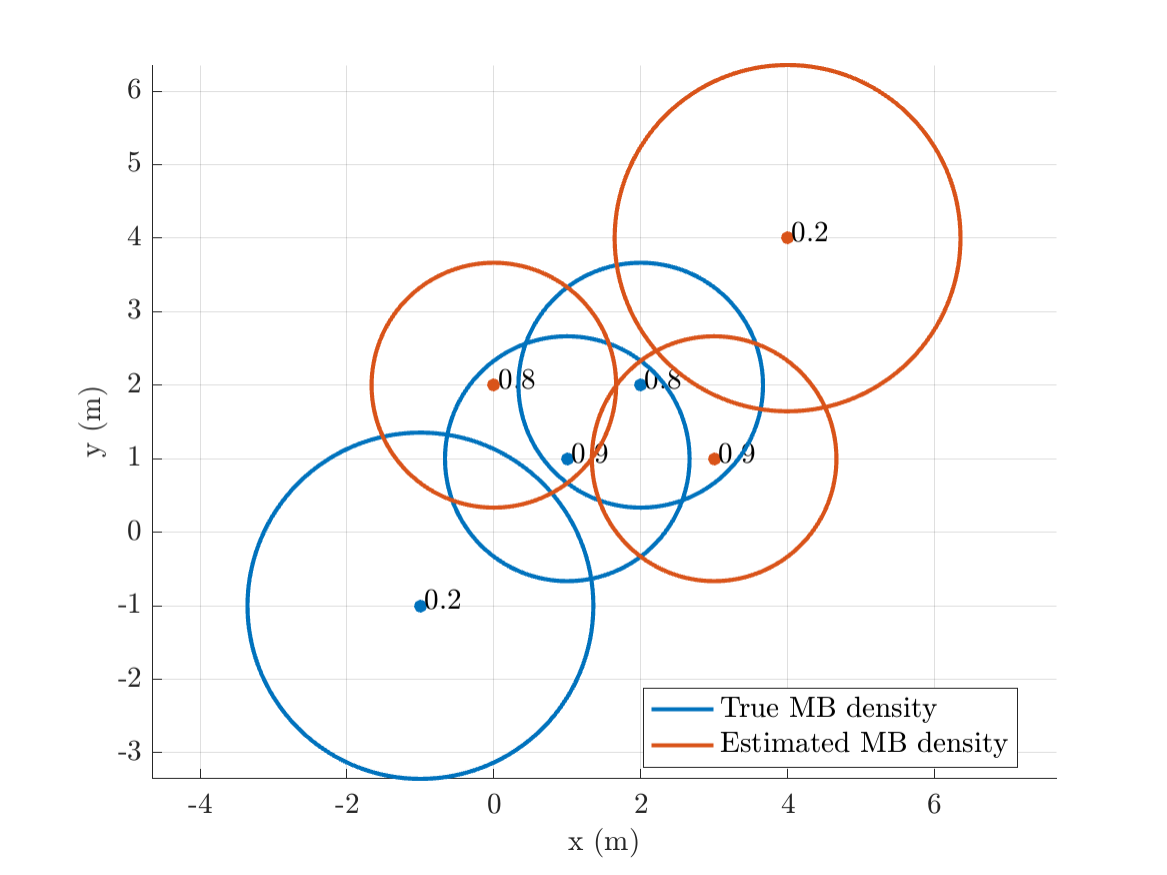}}
    \caption{Example 2: the true and estimated MB set density. Each Bernoulli density has Gaussian single-object density, and its existence probability is shown next to its Gaussian mean.}
    \label{fig_ex3}
\end{figure}

\begin{figure}[!t]
    \centerline{\includegraphics[width=\linewidth]{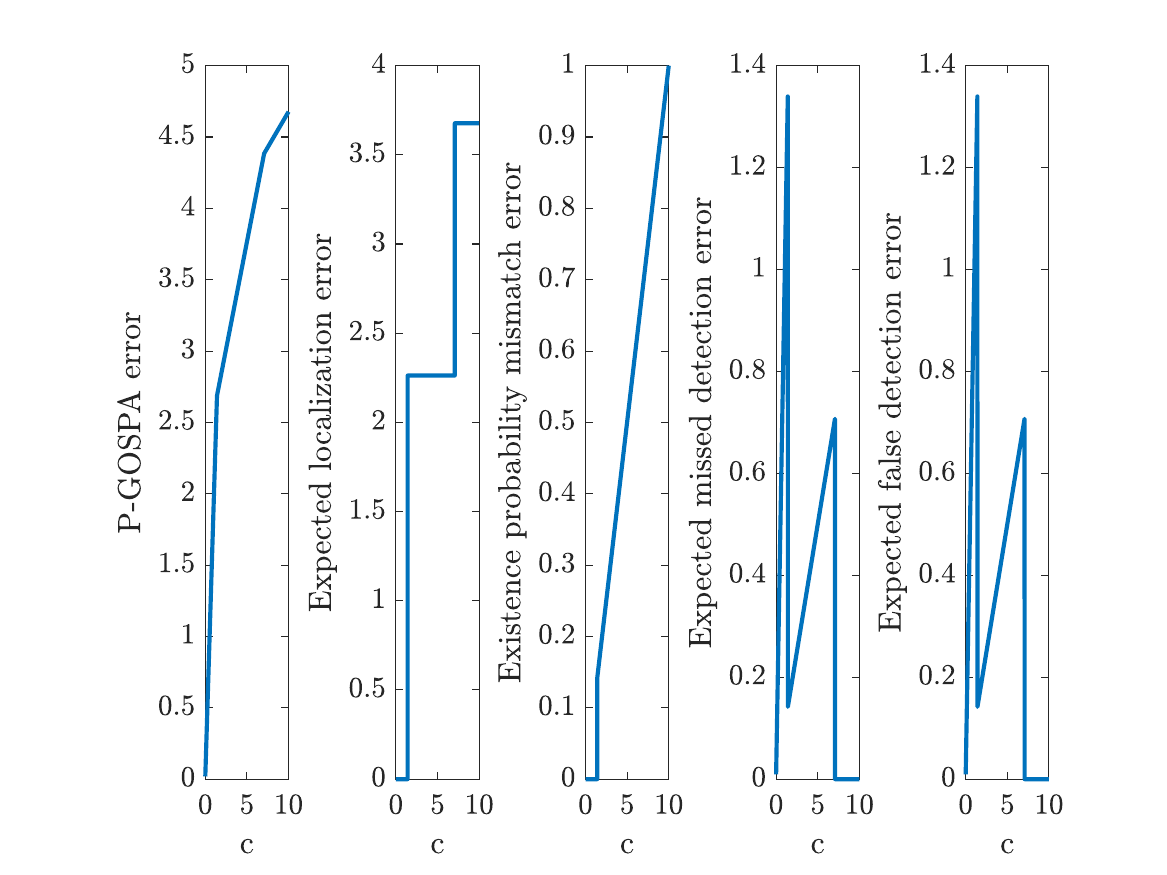}}
    \caption{Example 2: P-GOSPA and its decomposition versus $c$.}
    \label{fig_ex4}
\end{figure}

We consider a 2D example, in which both the ground truth and the estimated MB densities have three Bernoulli components, as illustrated in Fig. \ref{fig_ex3}. In this example, we study how P-GOSPA and its decomposition change with varying cut-off distance $c$ (from 0.1 to 10 with grid size 0.1), which are shown in Fig. \ref{fig_ex4}. 

We will explain the results from the perspective of optimal assignment described in Proposition \ref{prop_pgospa_alpha2}. The behavior of P-GOSPA as $c$ changes can be thought of as gradually relaxing the assignment rules. Initially, when the cutoff distance $c$ is small, the matching process is highly restrictive, and most components remain unassigned. This leads to high missed detection and false detection errors because the algorithm cannot form associations. As $c$ increases and reaches certain thresholds, components that are closer to each other become eligible for assignment. This results in a sudden decrease in missed and false detection errors, but simultaneously introduces localization and existence probability mismatch errors for the assigned components. These errors reflect how accurately the associated components match in terms of their locations and probabilities. At higher values of $c$, all components eventually get assigned, eliminating missed and false detection errors entirely, while the remaining errors stabilize, showing that all objects are accounted for. This dynamic demonstrates how P-GOSPA captures different aspects of error depending on the assignment flexibility dictated by $c$.

\begin{figure}[!t]
    \center
    \includegraphics[width=\linewidth]{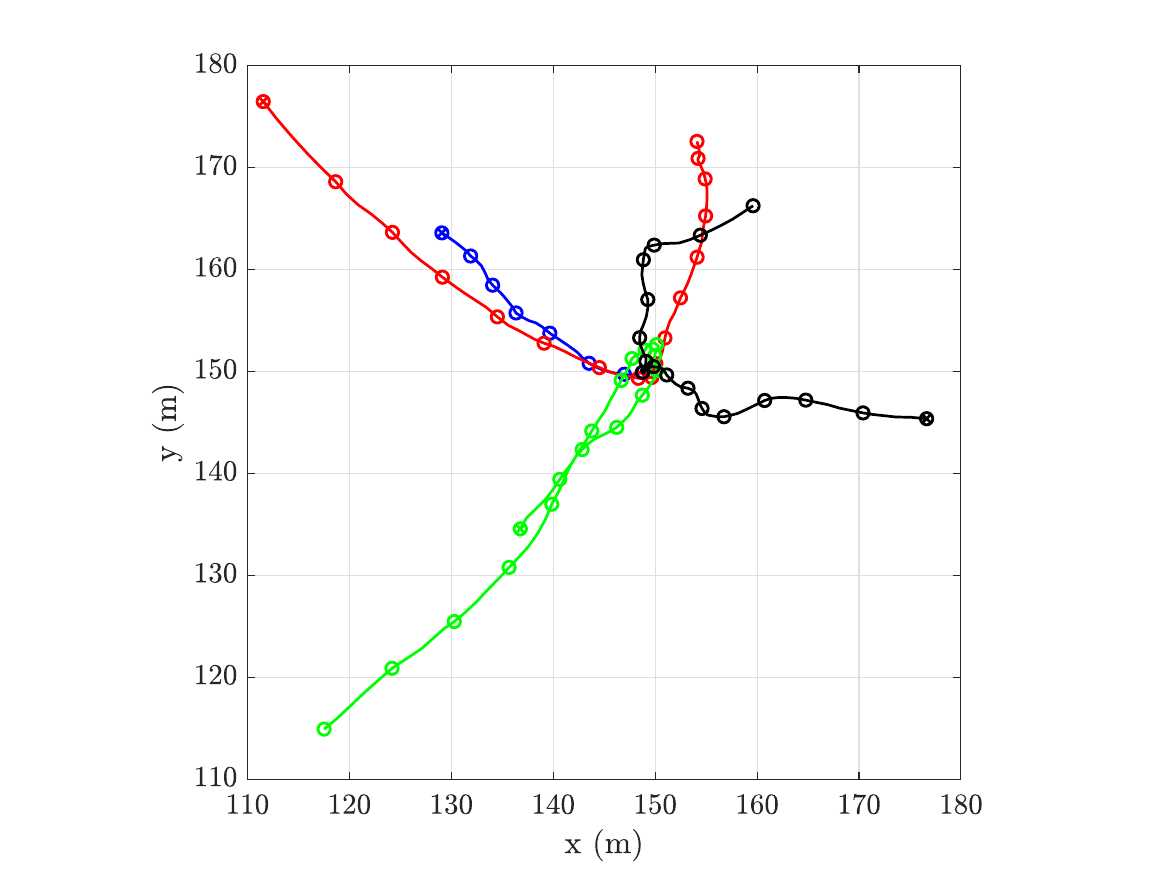}
    \includegraphics[width=\linewidth]{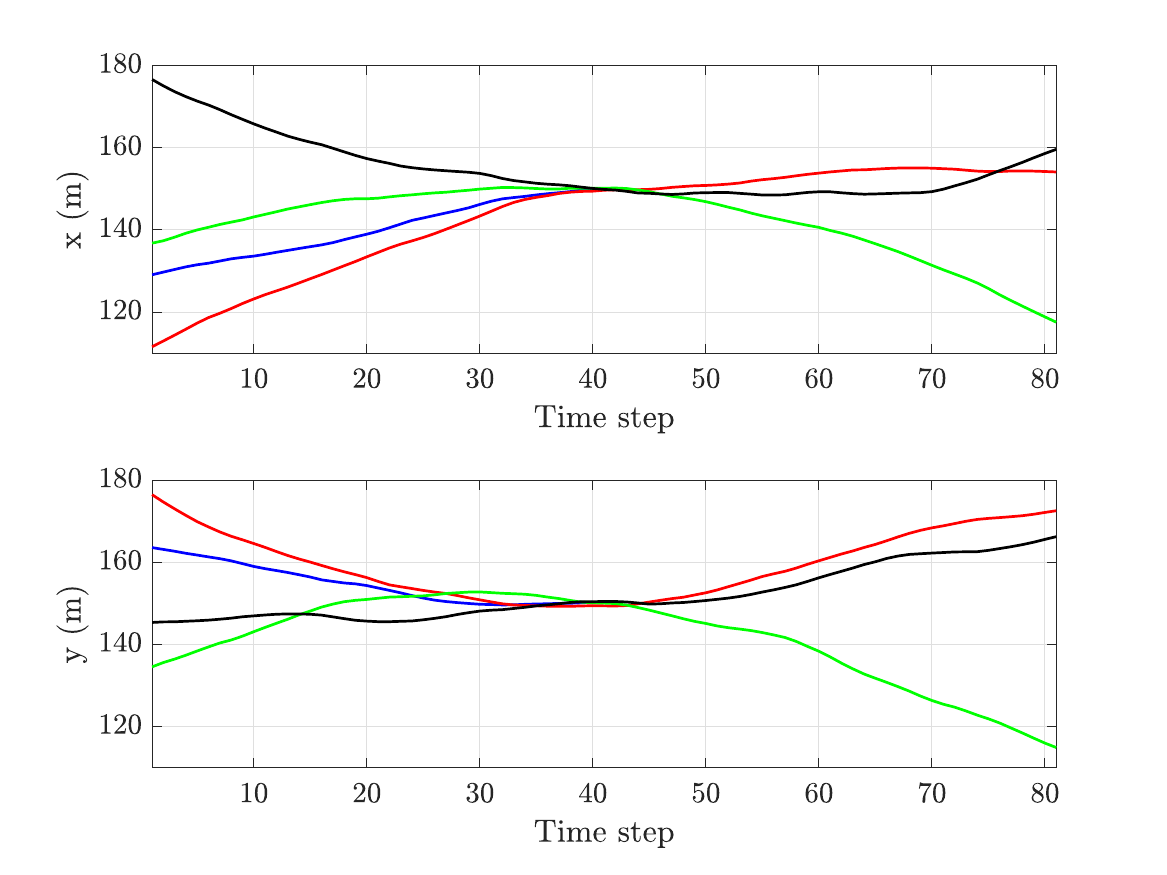}
    \caption{Ground truth object trajectories used in the simulation. The figure at the top shows the two-dimensional trajectories, and how their decompositions into x and y dimensions vary over time is illustrated in the two figures at the bottom. There are four objects, and they move in proximity at time step 40 \cite{garcia2018poisson}. Three objects remain present throughout the simulation, while one object disappears at time step 40. Object positions are marked with circle every 5 time steps, and the initial positions are also marked with cross.}
    \label{fig_gt}
\end{figure}

\begin{figure}[!t]
    \center
    \includegraphics[width=0.7\linewidth]{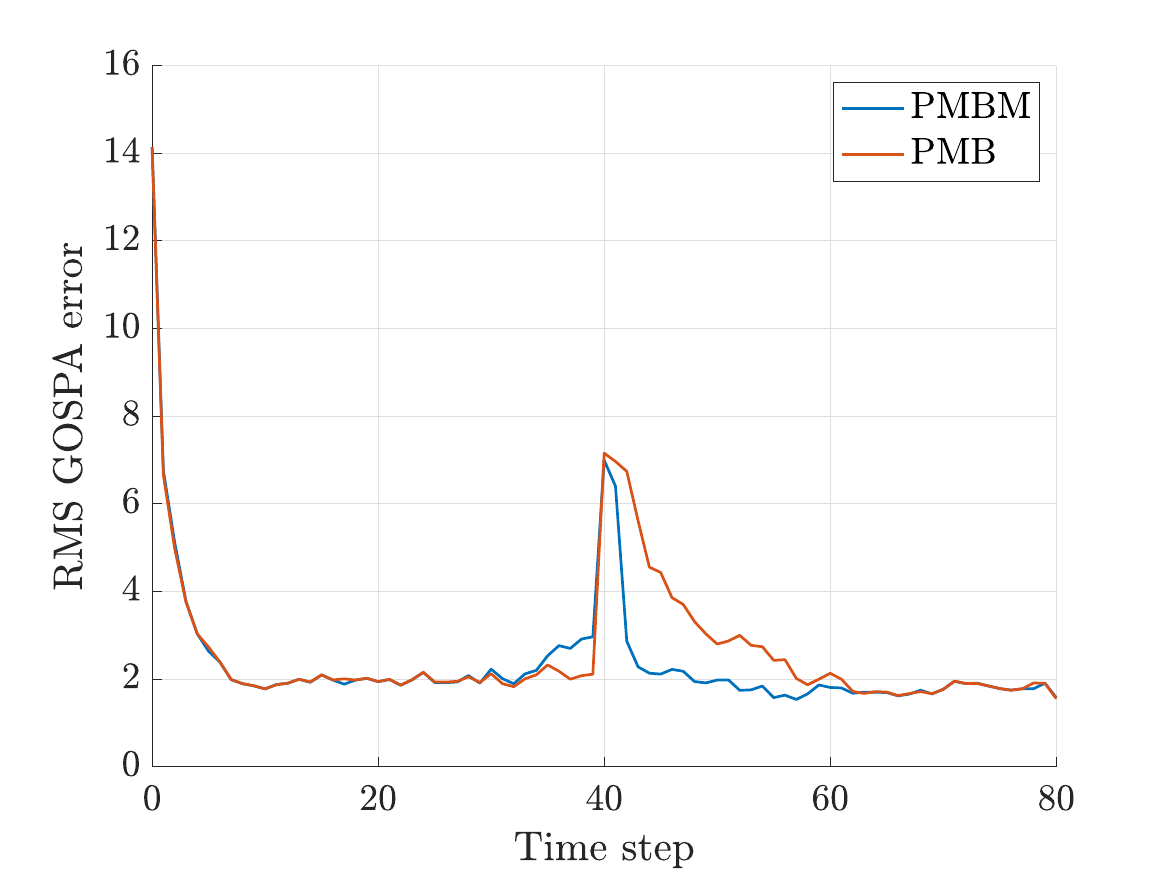}
    \includegraphics[width=0.7\linewidth]{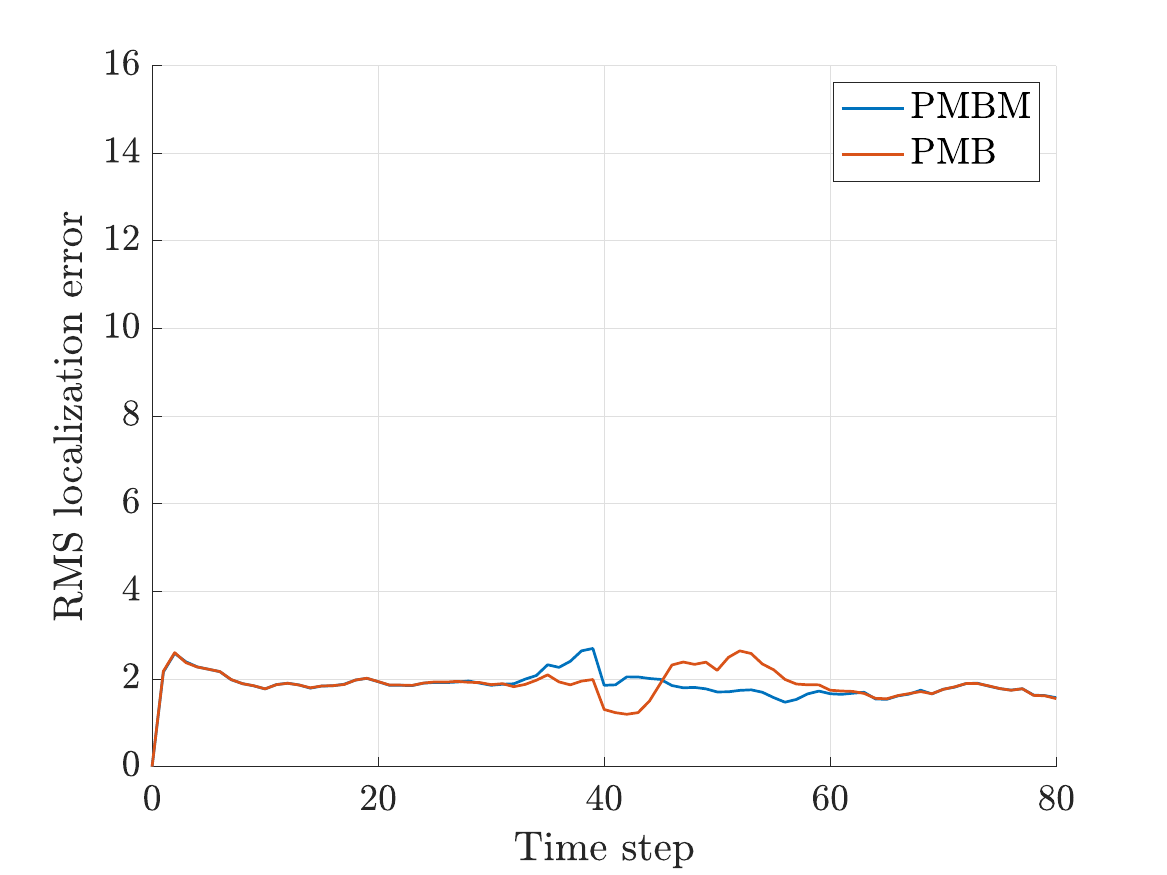}
    \includegraphics[width=0.7\linewidth]{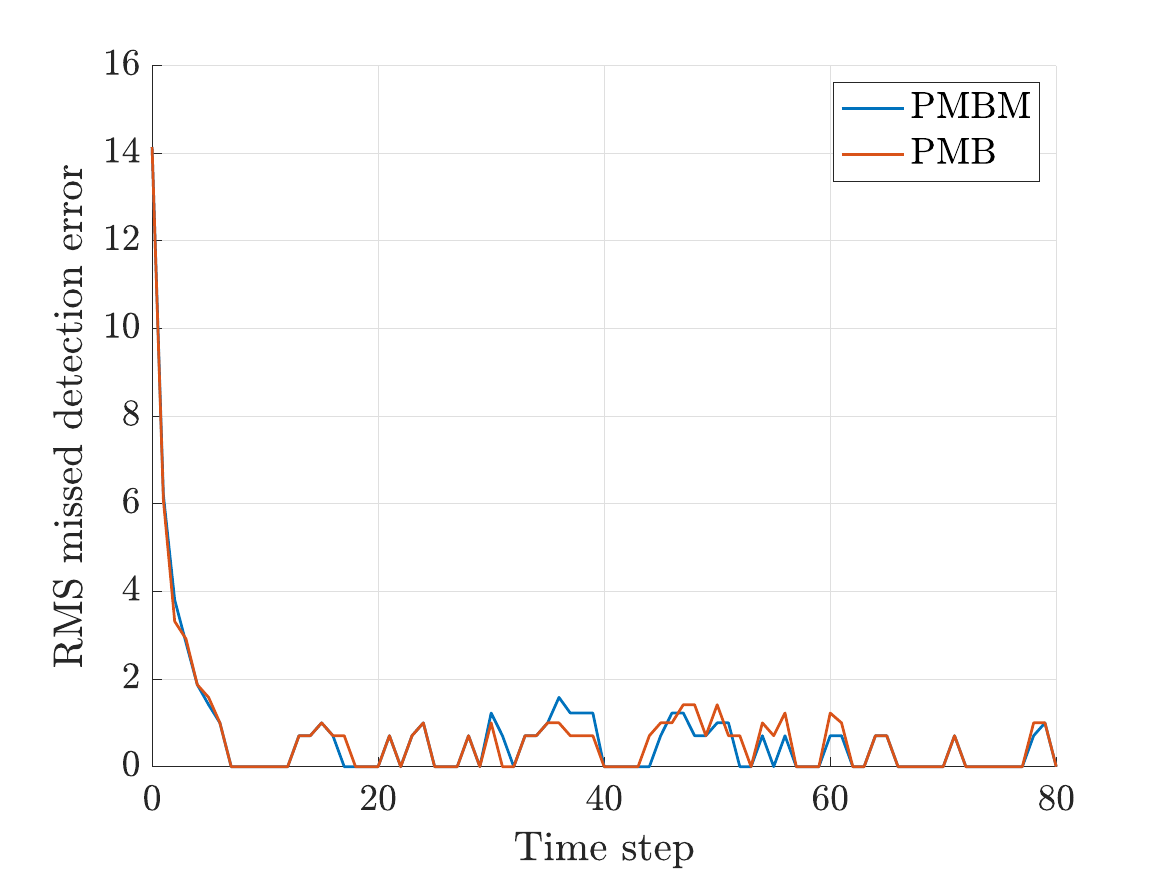}
    \includegraphics[width=0.7\linewidth]{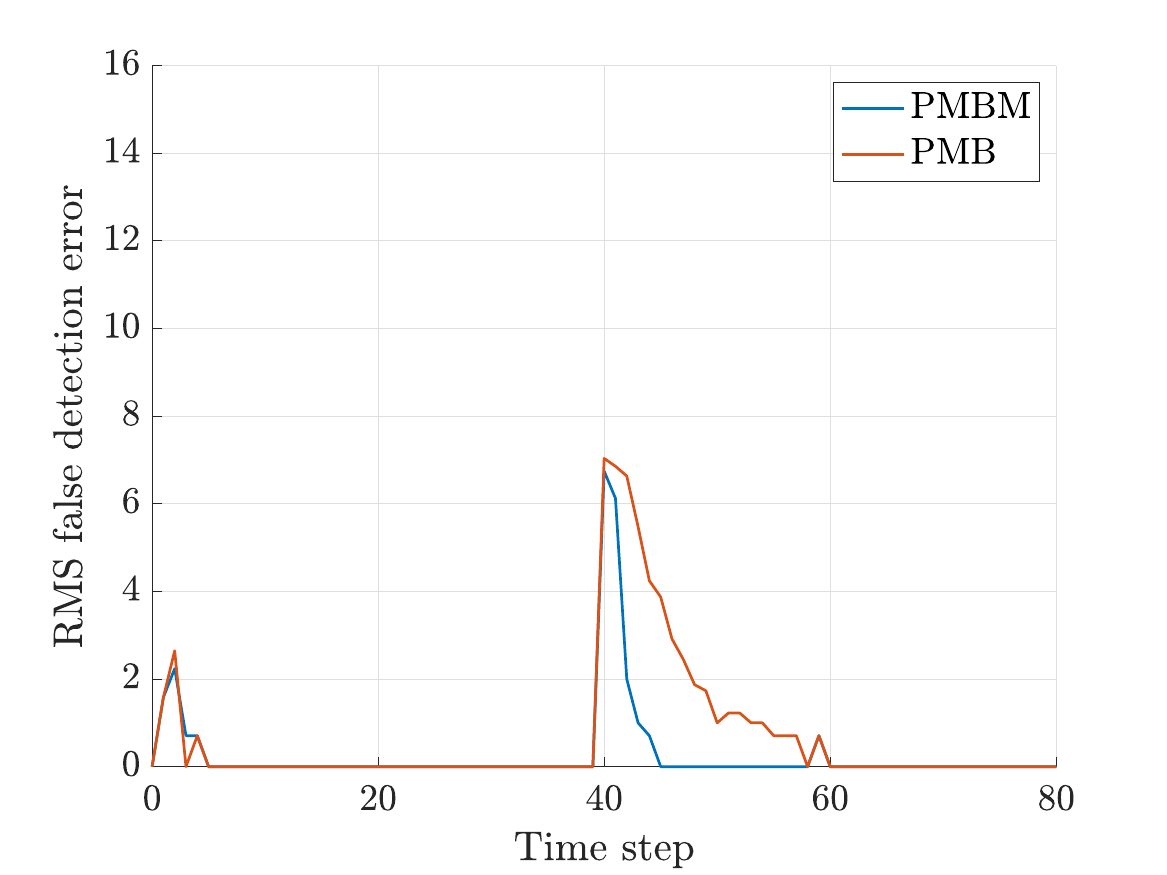}
    \caption{RMS GOSPA and its decomposition over time.}
    \label{fig_gospa}
\end{figure}

\begin{figure}[!t]
    \center
    \includegraphics[width=0.67\linewidth]{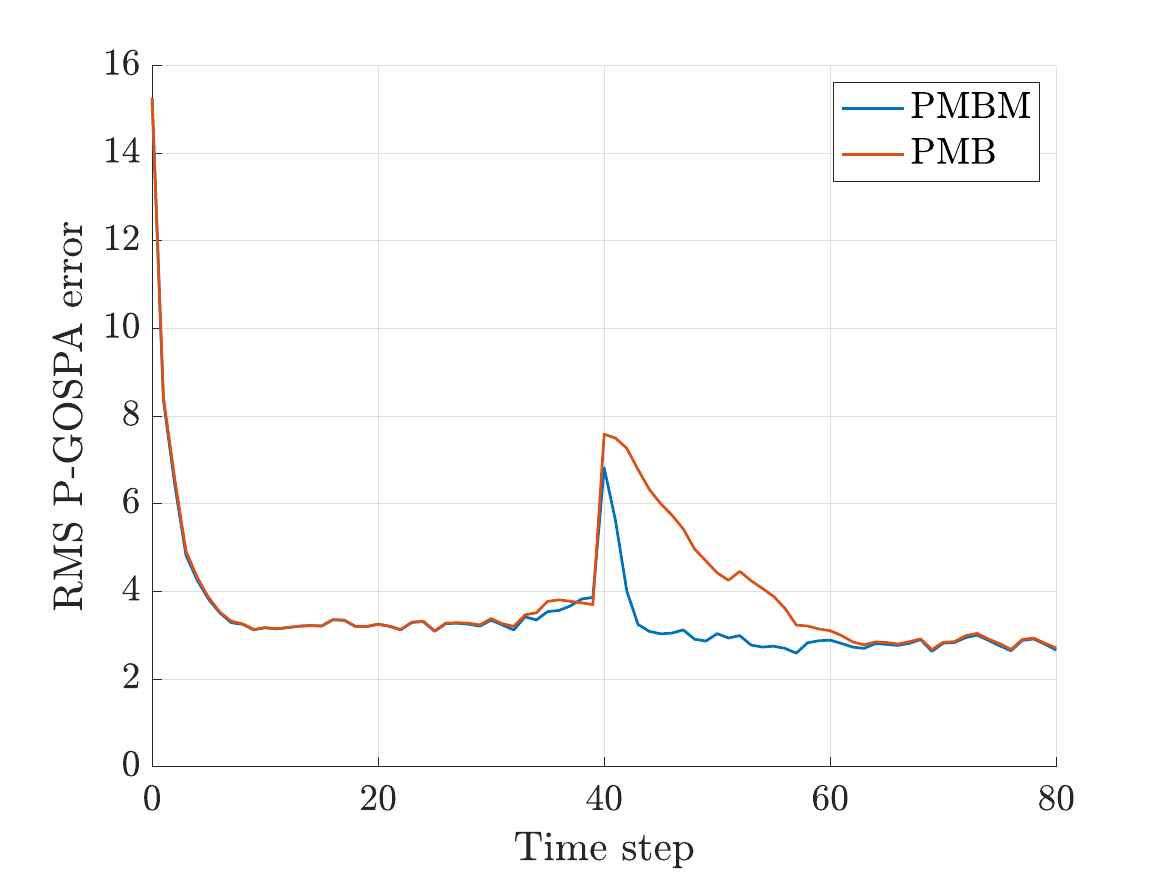}
    \includegraphics[width=0.67\linewidth]{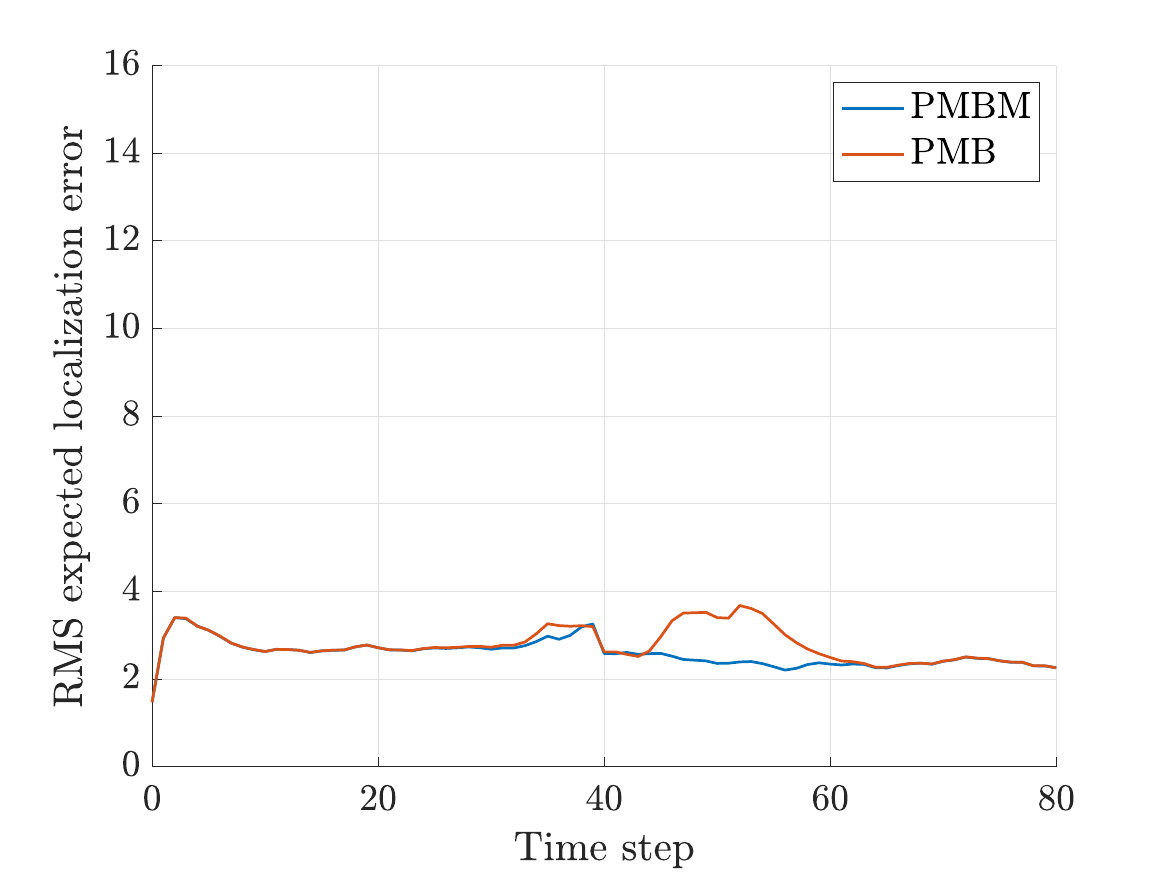}
    \includegraphics[width=0.67\linewidth]{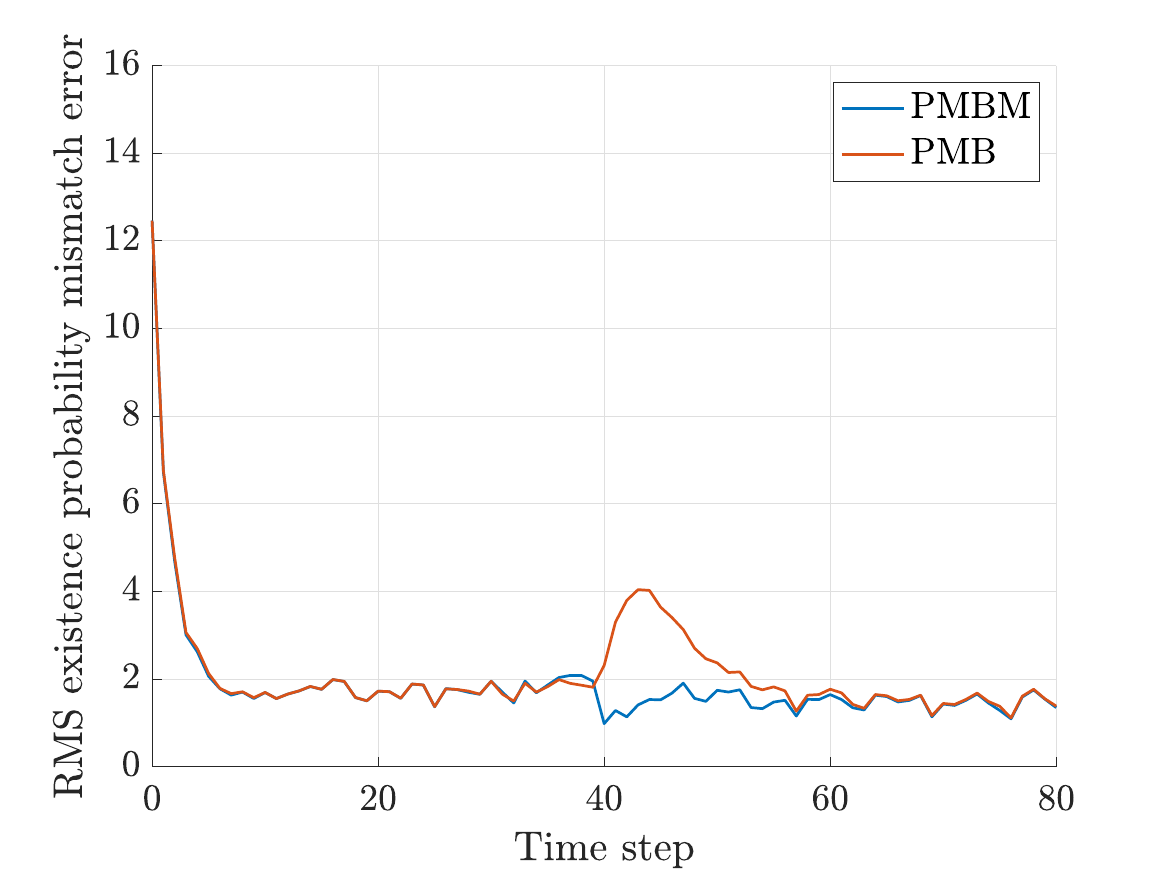}
    \includegraphics[width=0.67\linewidth]{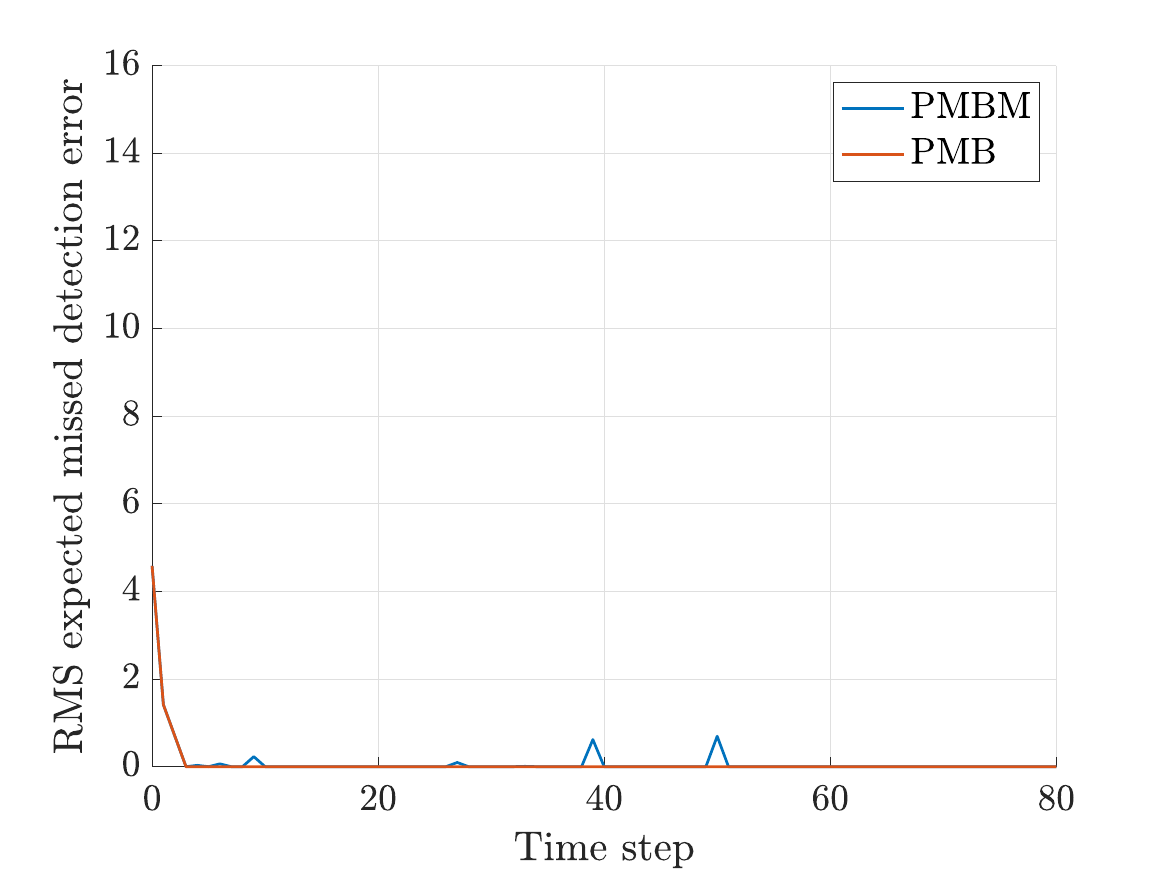}
    \includegraphics[width=0.67\linewidth]{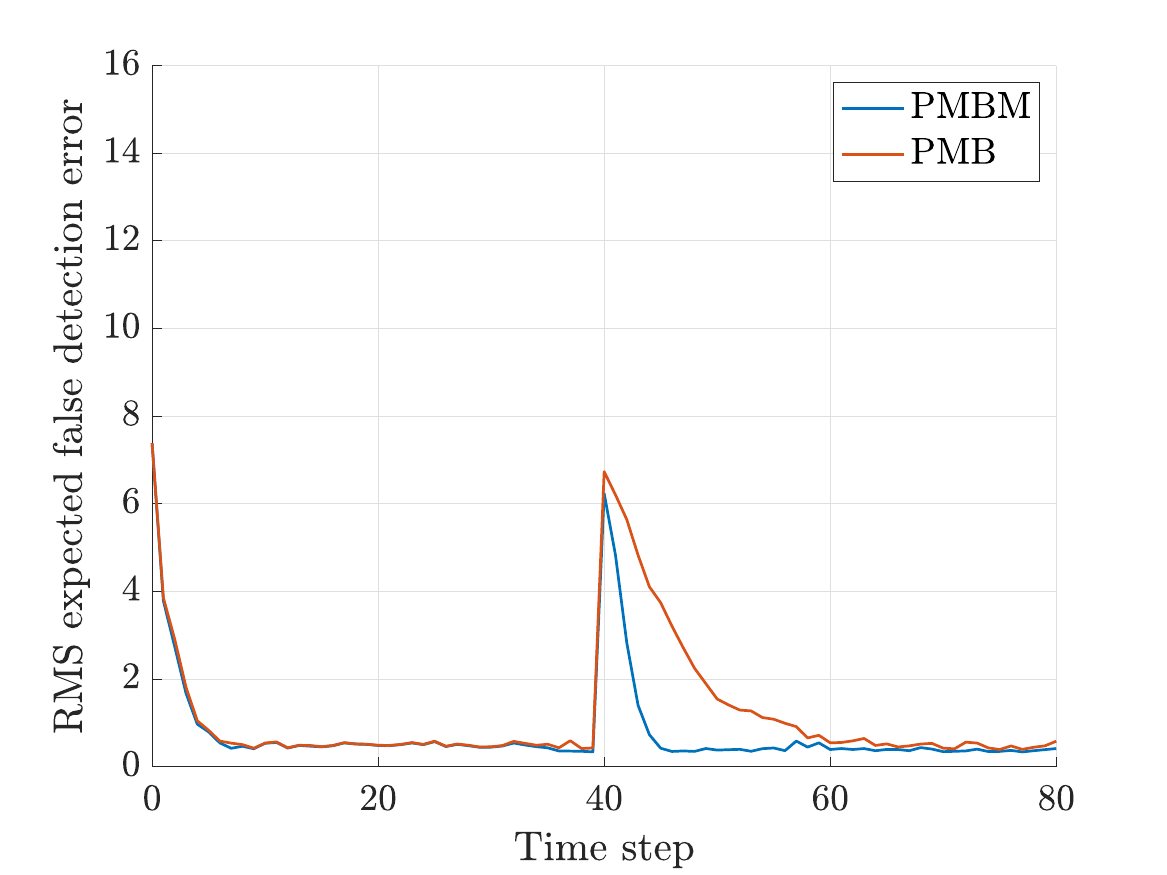}
    \caption{RMS P-GOSPA and its decomposition over time.}
    \label{fig_pgospa}
\end{figure}

\begin{figure}
    \center
    \includegraphics[width=\linewidth]{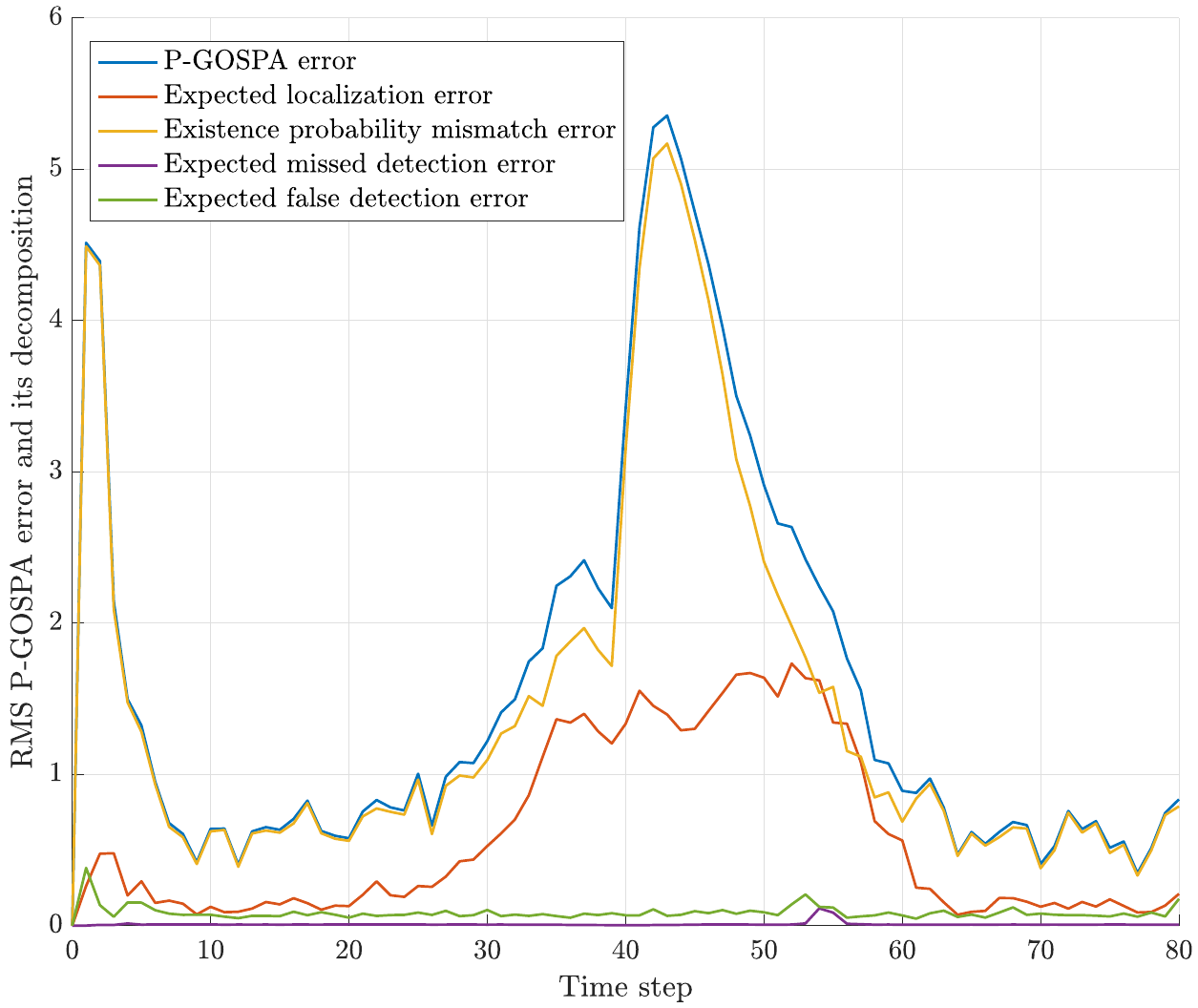}
    \caption{RMS P-GOSPA and its decomposition over time, where P-GOSPA is calculated between the MBM of the updated PMBM density (prior to the MB approximation in PMB filtering) and the MB density (after the approximation).}
    \label{fig_pgospa_pmb}
\end{figure}

\section{Simulation Results}

In this section, we compare the GOSPA and P-GOSPA metrics by using them to evaluate the multi-object filtering performance of the Poisson multi-Bernoulli mixture (PMBM) filter \cite{garcia2018poisson} and the track-oriented Poisson multi-Bernoulli (PMB) filter \cite{williams2015marginal}. The closed-form solution to MOT for the standard multi-object models with Poisson birth \cite{mahler2014advances} is given by the PMBM filter \cite{williams2015marginal,garcia2018poisson}. The multi-object posterior density of the PMBM filter follows a PMBM form, where the set of undetected objects is represented by a Poisson point process and the set of detected objects is represented by an MB mixture (MBM). If we merge the MBM into a single MB after every update step in a track-oriented fashion, we obtain a PMB filter, which is an efficient approximation of the PMBM filter.

Both filters are implemented with the following parameters: ellipsoidal gating size 20, maximum number of global hypotheses 200 (found using Murty's algorithm \cite{crouse2016implementing}), threshold for pruning the Poisson intensity weights $10^{-5}$, threshold for pruning Bernoulli components $10^{-5}$. In addition, for the PMBM filter, the threshold for pruning global hypotheses is $10^{-4}$. For PMBM, to fully quantify the uncertainty information in its MBM, we first compute P-GOSPA between each MB density and the set of ground truth objects and then take the weighted sum. Performance evaluation using GOSPA requires an estimator. For both filters, we report object position estimates from Bernoulli components with existence probabilities greater than 0.4, selecting them from the MB with the highest weight. In both GOSPA and P-GOSPA, we set $c = 10$ and $p = 2$.

In the simulation, we consider the same scenario as in \cite{garcia2018poisson} with an area $[\SI{0}{\meter},\SI{300}{\meter}]\times[\SI{0}{\meter},\SI{300}{\meter}]$. Object states include 2D position and velocity and are generated according to a Poisson point process birth model with an intensity of $0.005$. The birth density follows a single Gaussian density with a mean of $[100,0,100,0]^T$ and a covariance matrix given by $\text{diag}([150^2,1,150^2,1])$. We use the nearly constant velocity motion model with a sampling period of $\SI{1}{\second}$ and a noise standard deviation of 0.1. The measurement model follows a linear Gaussian distribution with an identity noise covariance matrix. We also use object survival probability 0.99, detection probability 0.9, and Poisson clutter with Poisson rate 10 and uniform density. The ground truth object trajectories are illustrated in Fig. \ref{fig_gt}.

We conduct 100 Monte Carlo simulations, and compute the root-mean-square (RMS) GOSPA and P-GOSPA errors and their decomposition at each time step for each filter. The GOSPA error and its decomposition over time are shown in Fig. \ref{fig_gospa}, whereas the P-GOSPA error and its decomposition over time are shown in Fig. \ref{fig_pgospa}. The results show that GOSPA and P-GOSPA trends generally align, both increasing as objects move in close proximity, with a notable change in false detection error when one object disappears at time step 40. One noticeable difference is that PMBM shows larger GOSPA error than PMB when objects move closer before time step 40. As a comparison, PMBM almost consistently outperforms PMB in terms of P-GOSPA. The underlying reason for this difference becomes apparent when comparing the figures illustrating the RMS (expected) localization errors. Specifically, the means of Gaussian object densities reported by PMB are generally closer to the ground truth object states, but they also have larger covariance.

In addition to performance evaluation, P-GOSPA can be also used to quantify the MB approximation error in PMB filtering. Fig. \ref{fig_pgospa_pmb} shows the P-GOSPA error calculated between the MBM of the updated PMBM density (prior to the MB approximation in PMB filtering) and the MB density (after the approximation). The results confirm that PMB is a less accurate representation of the multi-object posterior compared to PMBM, primarily due to its weaker handling of object cardinality uncertainties.

\section{Conclusions}

In this correspondence, we introduced a metric for evaluating the performance of multi-object filters with uncertainties, called P-GOSPA. P-GOSPA is a probabilistic generalization of the GOSPA metric, extending it to the space of MB densities. We demonstrated that, under specific parameter settings, P-GOSPA can be decomposed into four components: expected localization error and existence probability mismatch error for correctly detected objects, as well as expected missed detection and false detection errors. An interesting direction for future work would be to extend P-GOSPA to evaluate sets of trajectories \cite{garcia2020metric}.





\appendices

\section{Proof of the triangle inequality of P-GOSPA}\label{appendix_triangle_pgospa}

In the proof, we will make use of an extension of the Minkowski's inequality to sequences of different lengths~\cite[pp. 165]{kubrusly2011elements}. Specifically, for two sequences $(a_i)^{m}_{i=1}$ and $(b_i)^n_{i=1}$ with $m \leq n$, by appending the shorter sequence $(a_i)$ such that $a_i = 0$ for $i = m+1,\dots,n$ and applying Minkowski's inequality on this extended sequence, we get
\begin{align}
    &\left(\sum_{i=1}^m|a_i + b_i|^p + \sum_{i = m+1}^n |b_i|^p\right)^{\frac{1}{p}} \nonumber \\ 
    &\leq \left(\sum_{i=1}^m|a_i|^p\right)^{\frac{1}{p}} + \left(\sum_{i=1}^n|b_i|^p\right)^{\frac{1}{p}} \label{eq_Minkowski}
\end{align}
for $1 \leq p < +\infty$.

To prove that P-GOSPA satisfies the triangle inequality, we need to show that 
\begin{equation}
    d_p^{(c,\alpha)}(f_X,f_Y)  \leq d_p^{(c,\alpha)}(f_X,f_Z) + d_p^{(c,\alpha)}(f_Y,f_Z) \label{eq_triangleq}
\end{equation}
for any MB densities $f_X(\cdot)$, $f_Y(\cdot)$ and $f_Z(\cdot)$. 

\subsection{Triangle inequality for Bernoulli densities}

Before proceeding, we first consider the special case $n_X = n_Y = n_Z = 1$. In this case, P-GOSPA for Bernoulli set densities $f_X(\cdot)$ and $f_Y(\cdot)$ simplifies to
\begin{align}
    &d_p^{(c,\alpha)}(f_X,f_Y) \nonumber \\ &\triangleq \left(\min\left(r_x,r_y\right)d^{(c)}\left(p_x,p_y\right)^p + \left|r_x - r_y\right| \frac{c^p}{\alpha}\right)^{\frac{1}{p}}. \label{eq_bernoulli_pgospa}
\end{align}
To prove that \eqref{eq_bernoulli_pgospa} satisfies the triangle inequality, we also need to show that \eqref{eq_triangleq} holds.

Without loss of generality, we assume that $r_x \leq r_y$ as \eqref{eq_bernoulli_pgospa} is symmetric. The proof is divided into three cases, determined by the sizes of the existence probabilities $r_x$, $r_y$, and $r_z$. 

Case 1: $r_x \leq r_y \leq r_z$. It holds that
\begin{align}
    &d_p^{(c,\alpha)}(f_X,f_Y) \nonumber \\
    &=\left(r_xd^{(c)}\left(p_x,p_y\right)^p + \left|r_x - r_y\right|\frac{c^p}{\alpha} \right)^{\frac{1}{p}} \label{eq_case1_step_1} \\
    &\leq \left(\left|r_x - r_z\right|\frac{c^p}{\alpha} + \left|r_y - r_z\right|\frac{c^p}{\alpha} \right. \nonumber \\
    &~~~~~\left. + r_x\left(d^{(c)}\left(p_x,p_z\right) + d^{(c)}\left(p_y,p_z\right)\right)^p\right)^{\frac{1}{p}} \label{eq_case1_step_2} \\
    &= \left(\left|r_x - r_z\right|\frac{c^p}{\alpha} + \left|r_y - r_z\right|\frac{c^p}{\alpha} \right. \nonumber \\
    &~~~~~\left. + \left(r_x^{\frac{1}{p}}d^{(c)}\left(p_x,p_z\right) + r_x^{\frac{1}{p}}d^{(c)}\left(p_y,p_z\right)\right)^p\right)^{\frac{1}{p}} \label{eq_case1_step_3} \\
    &\leq \left(|r_x - r_z|\frac{c^p}{\alpha} + r_xd^{(c)}(p_x,p_z)^p\right)^{\frac{1}{p}} \nonumber \\
    &~~~+ \left(|r_y - r_z|\frac{c^p}{\alpha} + r_xd^{(c)}(p_y,p_z)^p\right)^{\frac{1}{p}} \label{eq_case1_step_4} \\
    &\leq \left(|r_x - r_z|\frac{c^p}{\alpha} + r_xd^{(c)}(p_x,p_z)^p\right)^{\frac{1}{p}} \nonumber \\
    &~~~+ \left(|r_y - r_z|\frac{c^p}{\alpha} + r_yd^{(c)}(p_y,p_z)^p\right)^{\frac{1}{p}} \label{eq_case1_step_5} \\
    &= d_p^{(c,\alpha)}(f_X,f_Z) + d_p^{(c,\alpha)}(f_Y,f_Z),
\end{align}
where we have first applied the triangle inequalities for $|r_x - r_y|$ and $d^{(c)}(p_x,p_y)$ from \eqref{eq_case1_step_1} to \eqref{eq_case1_step_2}, and then we have applied Minkowski's inequality from \eqref{eq_case1_step_3} to \eqref{eq_case1_step_4}.

Case 2: $r_x \leq r_z \leq r_y$. It holds that
\begin{align}
    &d_p^{(c,\alpha)}(f_X,f_Y) \nonumber \\
    &\leq \left(|r_x - r_z|\frac{c^p}{\alpha} + r_xd^{(c)}(p_x,p_z)^p\right)^{\frac{1}{p}} \nonumber \\
    &~~~+ \left(|r_y - r_z|\frac{c^p}{\alpha} + r_xd^{(c)}(p_y,p_z)^p\right)^{\frac{1}{p}} \label{eq_case2_step_4} \\
    &\leq \left(|r_x - r_z|\frac{c^p}{\alpha} + r_xd^{(c)}(p_x,p_z)^p\right)^{\frac{1}{p}} \nonumber \\
    &~~~+ \left(|r_y - r_z|\frac{c^p}{\alpha} + r_zd^{(c)}(p_y,p_z)^p\right)^{\frac{1}{p}} \label{eq_case2_step_5} \\
    &= d_p^{(c,\alpha)}(f_X,f_Z) + d_p^{(c,\alpha)}(f_Y,f_Z),
\end{align}
where the derivation is similar to Case 1.

Case 3: $r_z \leq r_x \leq r_y$. It holds that
\begin{align}
    &d_p^{(c,\alpha)}(f_X,f_Y) \nonumber \\
    &=\left(r_xd^{(c)}\left(p_x,p_y\right)^p + (r_y - r_x)\frac{c^p}{\alpha} \right)^{\frac{1}{p}} \label{eq_case3_step_1} \\
    &\leq \left(r_xd^{(c)}\left(p_x,p_y\right)^p + (r_y - r_x)\frac{c^p}{\alpha} \right. \nonumber \\
    &~~~~~\left. - (r_z - r_x)\left(\frac{2c^p}{\alpha} - d^{(c)}(p_x,p_y)^p\right) \right)^{\frac{1}{p}} \label{eq_case3_step_2} \\
    &= \left( r_zd^{(c)}(p_x,p_y)^p + (r_x + r_y - 2r_z)\frac{c^p}{\alpha}\right)^{\frac{1}{p}} \label{eq_case3_step_3}\\
    &\leq \left( r_z\left(d^{(c)}(p_x,p_z) + d^{(c)}(p_y,p_z)\right)^p \right. \nonumber \\
    &~~~~~\left. + (r_x + r_y - 2r_z)\frac{c^p}{\alpha}\right)^{\frac{1}{p}} \label{eq_case3_step_4}\\
    &= \left( \left(r_z^{\frac{1}{p}}d^{(c)}(p_x,p_z) + r_z^{\frac{1}{p}}d^{(c)}(p_y,p_z)\right)^p \right. \nonumber \\
    &~~~~~\left. + (r_x - r_z)\frac{c^p}{\alpha} + (r_y - r_z)\frac{c^p}{\alpha}\right)^{\frac{1}{p}} \label{eq_case3_step_5} \\
    &\leq \left(r_zd^{(c)}(p_x,p_z)^p + (r_x - r_z)\frac{c^p}{\alpha}\right)^{\frac{1}{p}} \nonumber \\
    &~~~+ \left(r_zd^{(c)}(p_y,p_z)^p + (r_y - r_z)\frac{c^p}{\alpha}\right)^{\frac{1}{p}} \label{eq_case3_step_6} \\
    &= d_p^{(c,\alpha)}(f_X,f_Z) + d_p^{(c,\alpha)}(f_Y,f_Z),
\end{align}
where we have applied the fact that $0 < \alpha \leq 2$ from \eqref{eq_case3_step_1} to \eqref{eq_case3_step_2}, the triangle inequality from \eqref{eq_case3_step_3} to \eqref{eq_case3_step_4}, and the Minkowski's inequality from \eqref{eq_case3_step_5} to \eqref{eq_case3_step_6}.

\subsection{Triangle inequality for multi-Bernoulli densities}

We have now finished the proof that P-GOSPA \eqref{eq_bernoulli_pgospa}  between two Bernoulli set densities satisfies the triangle inequality. We proceed to prove that P-GOSPA satisfies the triangle inequality for the general case \eqref{eq_pgospa}. Without loss of generality, we assume that $n_X \leq n_Y$ as P-GOSPA is symmetric. The proof is divided into three cases based on the number of Bernoulli components $n_X$, $n_Y$, and $n_Z$. 

Case 1: $n_X \leq n_Y \leq n_Z$. For any $\pi \in \Pi_{n_Y}$, we have 
\begin{align}
    &d_p^{(c,\alpha)}(f_X,f_Y) \nonumber \\ &\leq \left(\sum_{i=1}^{n_X}\left[\min\left(r_x^i,r_y^{\pi(i)}\right)d^{(c)}\left(p^i_x,p_y^{\pi(i)}\right)^p \right.\right. \nonumber \\
    &~~~~~\left.\left. + \left|r_x^i - r_y^{\pi(i)}\right|\frac{c^p}{\alpha}\right] + \frac{c^p}{\alpha}\sum_{i = n_X + 1}^{n_Y}r_y^{\pi(i)}\right)^{\frac{1}{p}}.
\end{align}
Using the triangle inequality on Bernoulli densities \eqref{eq_bernoulli_pgospa}, we have that for any $\pi \in \Pi_{n_Y}$ and $\sigma \in \Pi_{n_Z}$,
\begin{align}
    &d_p^{(c,\alpha)}(f_X,f_Y) \nonumber \\ 
    &\leq \left(\sum_{i=1}^{n_X}\left[\min\left(r_x^i,r_z^{\sigma(i)}\right)d^{(c)}\left(p^i_x,p_z^{\sigma(i)}\right)^p \right.\right. \nonumber \\
    &~~~~~\left. + \left|r_x^i - r_z^{\sigma(i)}\right|\frac{c^p}{\alpha}\right] + \sum_{i=1}^{n_X}\left[\min\left(r_y^{\pi(i)},r_z^{\sigma(i)}\right)\right.\nonumber \\
    &~~~~~\times d^{(c)}\left(p^{\pi(i)}_y,p_z^{\sigma(i)}\right)^p + \left.\left|r_y^{\pi(i)} - r_z^{\sigma(i)}\right|\frac{c^p}{\alpha}\right] \nonumber \\
    &~~~~~+\left.\frac{c^p}{\alpha}\sum_{i = n_X + 1}^{n_Y}r_y^{\pi(i)}\right)^{\frac{1}{p}}. \\
    &\leq \left(\sum_{i=1}^{n_X}\left[\min\left(r_x^i,r_z^{\sigma(i)}\right)d^{(c)}\left(p^i_x,p_z^{\sigma(i)}\right)^p \right.\right. \nonumber \\
    &~~~~~\left. + \left|r_x^i - r_z^{\sigma(i)}\right|\frac{c^p}{\alpha}\right] + \sum_{i=1}^{n_X}\left[\min\left(r_y^{\pi(i)},r_z^{\sigma(i)}\right)\right.\nonumber \\
    &~~~~~\times d^{(c)}\left(p^{\pi(i)}_y,p_z^{\sigma(i)}\right)^p + \left.\left|r_y^{\pi(i)} - r_z^{\sigma(i)}\right|\frac{c^p}{\alpha}\right] \nonumber \\
    &~~~~~+\left.\frac{c^p}{\alpha}\sum_{i = n_X + 1}^{n_Y} \left|r_y^{\pi(i)} - r_z^{\sigma(i)}\right| + \frac{c^p}{\alpha}\sum_{i = n_X+1}^{n_Y}r_z^{\sigma(i)}\right)^{\frac{1}{p}} \label{case1_step_2} \\
    &\leq \left(\sum_{i=1}^{n_X}\left[\min\left(r_x^i,r_z^{\sigma(i)}\right)d^{(c)}\left(p^i_x,p_z^{\sigma(i)}\right)^p \right.\right. \nonumber \\
    &~~~~~\left. + \left|r_x^i - r_z^{\sigma(i)}\right|\frac{c^p}{\alpha}\right] + \sum_{i=1}^{n_Y}\left[\min\left(r_y^{\pi(i)},r_z^{\sigma(i)}\right)\right.\nonumber \\
    &~~~~~\times d^{(c)}\left(p^{\pi(i)}_y,p_z^{\sigma(i)}\right)^p + \left.\left|r_y^{\pi(i)} - r_z^{\sigma(i)}\right|\frac{c^p}{\alpha}\right] \nonumber \\
    &~~~~~+\left. \sum_{i = n_X+1}^{n_Z}r_z^{\sigma(i)} + \frac{c^p}{\alpha}\sum_{i = n_Y+1}^{n_Z}r_z^{\sigma(i)}\right)^{\frac{1}{p}} \\
    &\leq \left(\sum_{i=1}^{n_X}\left[\min\left(r_x^i,r_z^{\sigma(i)}\right)d^{(c)}\left(p^i_x,p_z^{\sigma(i)}\right)^p \right.\right. \nonumber \\
    &~~~~~\left. + \left.\left|r_x^i - r_z^{\sigma(i)}\right|\frac{c^p}{\alpha}\right] + \frac{c^p}{\alpha}\sum_{i = n_X+1}^{n_Z}r_z^{\sigma(i)} \right)^{\frac{1}{p}} \nonumber \\
    &~~~+ \left(\sum_{i=1}^{n_Y}\left[\min\left(r_y^{\pi(i)},r_z^{\sigma(i)}\right)d^{(c)}\left(p^{\pi(i)}_y,p_z^{\sigma(i)}\right)^p \right.\right. \nonumber \\
    &~~~~~~~~\left. + \left.\left|r_y^{\pi(i)} - r_z^{\sigma(i)}\right|\frac{c^p}{\alpha}\right] + \frac{c^p}{\alpha}\sum_{i = n_Y+1}^{n_Z}r_z^{\sigma(i)} \right)^{\frac{1}{p}}, \label{case1_step_second_last}
\end{align}
where the Minkowski's inequality \eqref{eq_Minkowski} is applied to arrive at the last inequality. We further note that $\pi$ is a bijection that can be inverted. Denote the composition $\tau = \pi^{-1}\circ \sigma$, which is a permutation on $\{1,\dots,n_Z\}$. Then for any $\tau,\sigma \in \Pi_{n_Z}$, 
\begin{align}
    &d_p^{(c,\alpha)}(f_X,f_Y) \nonumber \\ 
    &\leq \left(\sum_{i=1}^{n_X}\left[\min\left(r_x^i,r_z^{\sigma(i)}\right)d^{(c)}\left(p^i_x,p_z^{\sigma(i)}\right)^p \right.\right. \nonumber \\
    &~~~~~\left. + \left.\left|r_x^i - r_z^{\sigma(i)}\right|\frac{c^p}{\alpha}\right] + \frac{c^p}{\alpha}\sum_{i = n_X+1}^{n_Z}r_z^{\sigma(i)} \right)^{\frac{1}{p}} \nonumber \\
    &~~~+ \left(\sum_{i=1}^{n_Y}\left[\min\left(r_y^{i},r_z^{\tau(i)}\right)d^{(c)}\left(p^{i}_y,p_z^{\tau(i)}\right)^p \right.\right. \nonumber \\
    &~~~~~~~~\left. + \left.\left|r_y^{i} - r_z^{\tau(i)}\right|\frac{c^p}{\alpha}\right] + \frac{c^p}{\alpha}\sum_{i = n_Y+1}^{n_Z}r_z^{\tau(i)} \right)^{\frac{1}{p}}, \label{case1_step_last}
\end{align}
which also holds for $\sigma$ and $\tau$ that minimize the first and second term on the right-hand side. 

This proves the triangle inequality for the case $n_X \leq n_Y \leq n_Z$.

Case 2: $n_X \leq n_Z \leq n_Y$. Similar to Case 1, for any $\pi \in \Pi_{n_Y}$ and $\sigma \in \Pi_{n_Z}$, we have 
\begin{align}
    &d_p^{(c,\alpha)}(f_X,f_Y) \nonumber \\ 
    &\leq \left(\sum_{i=1}^{n_X}\left[\min\left(r_x^i,r_z^{\sigma(i)}\right)d^{(c)}\left(p^i_x,p_z^{\sigma(i)}\right)^p \right.\right. \nonumber \\
    &~~~~~\left. + \left|r_x^i - r_z^{\sigma(i)}\right|\frac{c^p}{\alpha}\right] + \sum_{i=1}^{n_X}\left[\min\left(r_y^{\pi(i)},r_z^{\sigma(i)}\right)\right.\nonumber \\
    &~~~~~\times d^{(c)}\left(p^{\pi(i)}_y,p_z^{\sigma(i)}\right)^p + \left.\left|r_y^{\pi(i)} - r_z^{\sigma(i)}\right|\frac{c^p}{\alpha}\right] \nonumber \\
    &~~~~~+\left.\frac{c^p}{\alpha}\sum_{i = n_X + 1}^{n_Z} \left|r_y^{\pi(i)} - r_z^{\sigma(i)}\right| + \frac{c^p}{\alpha}\sum_{i = n_X+1}^{n_Z}r_z^{\sigma(i)}\right)^{\frac{1}{p}} \\
    &\leq \left(\sum_{i=1}^{n_X}\left[\min\left(r_x^i,r_z^{\sigma(i)}\right)d^{(c)}\left(p^i_x,p_z^{\sigma(i)}\right)^p \right.\right. \nonumber \\
    &~~~~~\left. + \left|r_x^i - r_z^{\sigma(i)}\right|\frac{c^p}{\alpha}\right] + \sum_{i=1}^{n_Z}\left[\min\left(r_y^{\pi(i)},r_z^{\sigma(i)}\right)\right.\nonumber \\
    &~~~~~\times d^{(c)}\left(p^{\pi(i)}_y,p_z^{\sigma(i)}\right)^p + \left.\left|r_y^{\pi(i)} - r_z^{\sigma(i)}\right|\frac{c^p}{\alpha}\right] \nonumber \\
    &~~~~~+\left. \frac{c^p}{\alpha}\sum_{i = n_X+1}^{n_Z}r_z^{\sigma(i)} + \frac{c^p}{\alpha}\sum_{i = n_Z+1}^{n_Y}r_y^{\pi(i)}\right)^{\frac{1}{p}} \\
    &\leq \left(\sum_{i=1}^{n_X}\left[\min\left(r_x^i,r_z^{\sigma(i)}\right)d^{(c)}\left(p^i_x,p_z^{\sigma(i)}\right)^p \right.\right. \nonumber \\
    &~~~~~\left. + \left.\left|r_x^i - r_z^{\sigma(i)}\right|\frac{c^p}{\alpha}\right] + \frac{c^p}{\alpha}\sum_{i = n_X+1}^{n_Z}r_z^{\sigma(i)} \right)^{\frac{1}{p}} \nonumber \\
    &~~~+ \left(\sum_{i=1}^{n_Z}\left[\min\left(r_y^{\pi(i)},r_z^{\sigma(i)}\right)d^{(c)}\left(p^{\pi(i)}_y,p_z^{\sigma(i)}\right)^p \right.\right. \nonumber \\
    &~~~~~~~~\left. + \left.\left|r_y^{\pi(i)} - r_z^{\sigma(i)}\right|\frac{c^p}{\alpha}\right] + \frac{c^p}{\alpha}\sum_{i = n_Z+1}^{n_Y}r_y^{\pi(i)} \right)^{\frac{1}{p}}.
\end{align}
The rest of the derivation is similar to the derivation from \eqref{case1_step_second_last} to \eqref{case1_step_last}.

This proves the triangle inequality for the case $n_X \leq n_Z \leq n_Y$.

Case 3: $n_Z \leq n_X \leq n_Y$. For any $\pi \in \Pi_{n_Y}$, we have 
\begin{align}
    &d_p^{(c,\alpha)}(f_X,f_Y) \nonumber \\ 
    &\leq \left(\sum_{i=1}^{n_X}\left[\min\left(r_x^i,r_y^{\pi(i)}\right)d^{(c)}\left(p^i_x,p_y^{\pi(i)}\right)^p \right.\right. \nonumber \\
    &~~~~~\left.\left. + \left|r_x^i - r_y^{\pi(i)}\right|\frac{c^p}{\alpha}\right] + \frac{c^p}{\alpha}\sum_{i = n_X + 1}^{n_Y}r_y^{\pi(i)}\right)^{\frac{1}{p}} \label{case3_step1}\\
    &\leq \left(\sum_{i=1}^{n_Z}\left[\min\left(r_x^i,r_y^{\pi(i)}\right)d^{(c)}\left(p^i_x,p_y^{\pi(i)}\right)^p \right.\right. \nonumber \\
    &~~~~~\left. + \left|r_x^i - r_y^{\pi(i)}\right|\frac{c^p}{\alpha}\right] + \frac{c^p}{\alpha}\sum_{i = n_X + 1}^{n_Y}r_y^{\pi(i)} \nonumber \\
    &~~~~~\left.+ \sum_{i=n_Z+1}^{n_X}\left[\min\left(r_x^i,r_y^{\pi(i)}\right)\frac{2c^p}{\alpha} + \left|r_x^i - r_y^{\pi(i)}\right|\frac{c^p}{\alpha}\right]\right)^{\frac{1}{p}} \label{case3_step2}\\
    &= \left(\sum_{i=1}^{n_Z}\left[\min\left(r_x^i,r_y^{\pi(i)}\right)d^{(c)}\left(p^i_x,p_y^{\pi(i)}\right)^p \right.\right. \nonumber \\
    &~~~~~\left. + \left|r_x^i - r_y^{\pi(i)}\right|\frac{c^p}{\alpha}\right] + \frac{c^p}{\alpha}\sum_{i = n_X + 1}^{n_Y}r_y^{\pi(i)} \nonumber \\
    &~~~~~\left.+ \frac{c^p}{\alpha}\sum_{i=n_Z+1}^{n_X}\left(r_x^i + r_y^{\pi(i)}\right)\right)^{\frac{1}{p}} \\
    &= \left(\sum_{i=1}^{n_Z}\left[\min\left(r_x^i,r_y^{\pi(i)}\right)d^{(c)}\left(p^i_x,p_y^{\pi(i)}\right)^p \right.\right. \nonumber\\ &~~~~~+ \left.\left|r_x^i - r_y^{\pi(i)}\right|\frac{c^p}{\alpha}\right] \nonumber \\
    &~~~~~\left.+ \frac{c^p}{\alpha}\left(\sum_{i = n_Z + 1}^{n_Y}r_y^{\pi(i)} + \sum_{i=n_Z+1}^{n_X} r_x^i \right)\right)^{\frac{1}{p}},
\end{align}
where we have applied the fact that $d^{(c)}(p_x,p_y) \leq c$ and that $0 < \alpha \leq 2$ from \eqref{case3_step1} to \eqref{case3_step2}. In addition, similar to Case 1 and Case 2, by first applying the triangle inequality of \eqref{eq_bernoulli_pgospa} and then the Minkowski's inequality, we obtain for any $\pi \in \Pi_{n_Y}$ and $\sigma \in \Pi_{n_Z}$,
\begin{align}
    &d_p^{(c,\alpha)}(f_X,f_Y) \nonumber \\ 
    &\leq \left(\sum_{i=1}^{n_Z}\left[\min\left(r_x^i,r_z^{\sigma(i)}\right)d^{(c)}\left(p^i_x,p_z^{\sigma(i)}\right)^p \right.\right. \nonumber \\
    &~~~~~\left. + \left.\left|r_x^i - r_z^{\sigma(i)}\right|\frac{c^p}{\alpha}\right] + \frac{c^p}{\alpha}\sum_{i = n_Z+1}^{n_X}r_x^{i} \right)^{\frac{1}{p}} \nonumber \\
    &~~~+ \left(\sum_{i=1}^{n_Z}\left[\min\left(r_y^{\pi(i)},r_z^{\sigma(i)}\right)d^{(c)}\left(p^{\pi(i)}_y,p_z^{\sigma(i)}\right)^p \right.\right. \nonumber \\
    &~~~~~~~~\left. + \left.\left|r_y^{\pi(i)} - r_z^{\sigma(i)}\right|\frac{c^p}{\alpha}\right] + \frac{c^p}{\alpha}\sum_{i = n_Z+1}^{n_Y}r_y^{\pi(i)} \right)^{\frac{1}{p}}.
\end{align}
The rest of the derivation is similar to the derivation from \eqref{case1_step_second_last} to \eqref{case1_step_last}.

This proves the triangle inequality for the case $n_Z \leq n_X \leq n_Y$.

\section{Proof of Proposition \ref{prop_Wasserstein_bernoulli}}\label{appendix_proposition_Wasserstein_bernoulli}

Suppose that we have two Bernoulli set densities $f_X(\cdot)$ and $f_Y(\cdot)$, defined on the space $\Omega = \{\emptyset\}\cup\{\{x\}\in\mathbb{R}^N\}$. Let $\mathcal{Q}(f_X,f_Y)$ denote the set of all the joint distributions $q$ for $(X,Y)$ that have marginals $f_X(\cdot)$ and $f_Y(\cdot)$, respectively. Following \cite{villani2009optimal}, the $p$-Wasserstein distance between $f_X(\cdot)$ and $f_Y(\cdot)$ can be defined as 
\begin{align}
    &W_p(f_X,f_Y)\nonumber\\ 
    &= \left(\text{inf}_{q\in\mathcal{Q}(f_X,f_Y)}\iint \bar{d}^{(c,\alpha)}_p(X,Y)^p q(X,Y) \delta X \delta Y  \right)^{1/p},
\end{align}
where $p\geq 1$, and $\bar{d}^{(c,\alpha)}_p(X,Y)$ is the GOSPA metric between two Bernoulli sets $X$ and $Y$, such that
\begin{equation}
    \bar{d}^{(c,\alpha)}_p(X,Y)^p = \begin{cases}
            \min(\bar{d}(x,y),c)^p & X= \{x\},Y = \{y\} \\
            c^p/\alpha & X = \emptyset,Y \neq \emptyset \\
            c^p/\alpha & X \neq \emptyset,Y = \emptyset \\
            0 & X = \emptyset,Y = \emptyset.
        \end{cases}
\end{equation}

Our goal is to find the joint distribution $q\in\mathcal{Q}(f_X,f_Y)$ (if exists) that minimizes $W^p_p(f_X,f_Y)$, and the objective function can be expressed as 
\begin{align}
    &\iint \bar{d}^{(c,\alpha)}_p(X,Y)^p q(X,Y) \delta X \delta Y \label{eq_objective}\\
    &= \iint \min(\bar{d}(x,y),c)^p q(\{x\},\{y\})dx dy \nonumber \\
    &~~~+ \frac{c^p}{\alpha} \left(\int q(\{x\},\emptyset)dx + \int q(\emptyset,\{y\})dy\right) \nonumber \\
    &= \iint \min(\bar{d}(x,y),c)^p p(x,y)dx dy \iint q(\{x\},\{y\}) dx dy \nonumber \\
    &~~~+ \frac{c^p}{\alpha} \left(\int q(\{x\},\emptyset)dx + \int q(\emptyset,\{y\})dy\right) \label{eq_objective_upperbound}
\end{align}
where in the second equality the joint density $p(x,y)$ is 
\begin{equation}
    p(x,y) = \frac{q(\{x\},\{y\})}{\iint q(\{x\},\{y\}) dx dy}.\label{eq_joint_density}
\end{equation}
The minimization of \eqref{eq_objective_upperbound} can be formulated as an optimal transport problem, and the objective is to find the optimal weights 
\begin{align}
    &\iint q(\{x\},\{y\})dx dy,\label{eq_weight1}\\
    &\int q(\{x\},\emptyset)dx + \int q(\emptyset,\{y\})dy,\label{eq_weight2}
\end{align}
and $q(\emptyset,\emptyset)$ (which has zero cost) that minimizes \eqref{eq_objective_upperbound}. We further assume that Bernoulli set densities $f_X(\cdot)$ and $f_Y(\cdot)$ are parameterized as in Lemma \ref{lemma_pgospa_bernoulli}. Then the weights are further subject to the marginal constraints
\begin{align}
    q(\emptyset,\emptyset) + \int q(\{x\},\emptyset) dx &= 1- r_y, \\
    q(\emptyset,\emptyset) + \int q(\emptyset,\{y\}) dy &= 1- r_x, \\
    \int q(\emptyset,\{y\}) dy + \iint q(\{x\},\{y\}) dx dy &= r_y, \\
    \int q(\{x\},\emptyset) dx + \iint q(\{x\},\{y\}) dx dy &= r_x.
\end{align}

We proceed to express \eqref{eq_objective_upperbound} in terms of $q(\emptyset,\emptyset)$, which yields
\begin{align}
    &\iint \min(\bar{d}(x,y),c)^p p(x,y)dx dy \nonumber\\
    &\times\left(r_x + r_y - 1 + q(\emptyset,\emptyset)\right)+\frac{c^p}{\alpha}\left(2-r_x-r_y-2q(\emptyset,\emptyset)\right). \label{eq_upperbound_2}
\end{align}
The derivative of \eqref{eq_upperbound_2} with respect to $q(\emptyset,\emptyset)$ is 
\begin{equation}
    \iint \min(\bar{d}(x,y),c)^p p(x,y)dx dy - \frac{2c^p}{\alpha},
\end{equation}
which is always no larger than zero since $0 < \alpha \leq 2$ and the first term is upper bounded by $c^p$. This means that we can always find a lower \eqref{eq_objective_upperbound} by increasing the weight $q(\emptyset,\emptyset)$. Since we have $\int q(\{x\},\emptyset) dx \geq 0$ and $\int q(\emptyset,\{y\}) dy \geq 0$, it holds that
\begin{equation}
    q(\emptyset,\emptyset) \leq \min(1-r_y,1-r_x) = 1 - \max(r_x,r_y).
\end{equation}
Then to minimize \eqref{eq_objective_upperbound}, we simply let 
\begin{equation}
    q(\emptyset,\emptyset) = 1 - \max(r_x,r_y).
\end{equation}
In this case, we also have
\begin{align}
    \iint q(\{x\},\{y\})dx dy &= \min(r_x,r_y),\label{eq_plugin_1}\\
    \int q(\{x\},\emptyset)dx + \int q(\emptyset,\{y\})dy &= |r_x - r_y|.\label{eq_plugin_2}
\end{align}

Note that these obtained optimal values are valid for any possible $p(x,y)$. Now we proceed to optimize over $p(x,y)$. For the first integral in the first term in \eqref{eq_objective_upperbound}, the following relation holds
\begin{align}
    &\text{inf}_{p(x,y)}\iint \min(\bar{d}(x,y),c)^p p(x,y)dx dy \nonumber\\
    &\leq \min\left(\text{inf}_{p(x,y)}\iint \bar{d}(x,y)^p p(x,y)dx dy,c^p\right), \label{eq_plugin_3}\\
    &= d^{(c)}(p_x,p_y)^p,\nonumber
\end{align}
where $d^{(c)}(p_x,p_y)$ is the cut-off metric of the Wasserstein distance between single-object densities $p_x(\cdot)$ and $p_y(\cdot)$ with $\bar{d}(x,y)$ as its cost function. Then, by plugging \eqref{eq_plugin_1} and \eqref{eq_plugin_2} into \eqref{eq_objective_upperbound} and using the inequality \eqref{eq_plugin_3}, we have
\begin{align}
    &\text{inf}_{q\in\mathcal{Q}(f_X,f_Y)}\iint \bar{d}^{(c,\alpha)}_p(X,Y)^p q(X,Y) \delta X \delta Y \nonumber\\
    &= \min(r_x,r_y)\text{inf}_{p(x,y)}\iint\min(\bar{d}(x,y),c)^p p(x,y)dx dy \nonumber\\
    &~~~+ |r_x - r_y|\frac{c^p}{\alpha}\nonumber \\
    &\leq \min(r_x,r_y)d^{(c)}(p_x,p_y)^p+ |r_x - r_y|\frac{c^p}{\alpha}.
\end{align}
Therefore, the P-GOSPA metric between two Bernoulli set densities \eqref{eq_bernoulli_cost} with the $p$-Wasserstein distance as the base distance can be interpreted as an upper bound on the $p$-Wasserstein distance using GOSPA as its cost function between the two Bernoulli set densities. 


This finishes the proof of Proposition \ref{prop_Wasserstein_bernoulli}.

\section{Proof of Lemma \ref{lemma_wasserstein_bernoulli}}\label{appendix_lemma_wasserstein_bernoulli}

Lemma \ref{lemma_wasserstein_bernoulli} is a special case of Proposition \ref{prop_Wasserstein_bernoulli}. 
Under the assumption that the single-object densities $p_x(\cdot)$ and $p_y(\cdot)$ are Dirac delta functions $\delta_x(\cdot)$ and $\delta_y(\cdot)$, their joint density $p(x,y)$ is a product of $\delta_x(\cdot)$ and $\delta_y(\cdot)$. Therefore, the inequality sign in \eqref{eq_plugin_3} to becomes an equality sign. 

\section{Proof of Proposition \ref{prop_pgospa_alpha2}}\label{appendix_proposition_pgospa_alpha2}

For $n_X \leq n_Y$ and $\alpha = 2$, we have 
\begin{align}
    &d_p^{(c,2)}(f_X,f_Y) \nonumber \\ 
    &= \left[\min_{\pi \in \Pi_{n_Y}}\left(\sum_{i=1}^{n_X}\left[\min\left(r_x^i,r_y^{\pi(i)}\right)d^{(c)}\left(p^i_x,p_y^{\pi(i)}\right)^p \right.\right.\right. \nonumber \\
    &~~~\left.\left.\left. + \left|r_x^i - r_y^{\pi(i)}\right|\frac{c^p}{2}\right] + \frac{c^p}{2}\sum_{i = n_X + 1}^{n_Y}r_y^{\pi(i)}\right)\right]^{\frac{1}{p}} \\
    &= \left[\min_{\pi \in \Pi_{n_Y}}\left(\sum_{i:d\left(p_x^i,p_y^{\pi(i)}\right)<c}\left[\min\left(r_x^i,r_y^{\pi(i)}\right)d\left(p^i_x,p_y^{\pi(i)}\right)^p \right.\right.\right. \nonumber \\
    &~~~\left. + \left|r_x^i - r_y^{\pi(i)}\right|\frac{c^p}{2}\right] + \frac{c^p}{2}\sum_{i = n_X + 1}^{n_Y}r_y^{\pi(i)} \nonumber \\ 
    &\left. +\left. \sum_{i:d\left(p_x^i,p_y^{\pi(i)}\right)\geq c} \left[\min\left(r_x^i,r_y^{\pi(i)}\right)c^p + \left|r_x^i - r_y^{\pi(i)}\right|\frac{c^p}{2}\right]\right)\right]^{\frac{1}{p}} \\
    &= \left[\min_{\pi \in \Pi_{n_Y}}\left(\sum_{i:d\left(p_x^i,p_y^{\pi(i)}\right)<c}\left[\min\left(r_x^i,r_y^{\pi(i)}\right)d\left(p^i_x,p_y^{\pi(i)}\right)^p \right.\right.\right. \nonumber \\
    &~~~\left. + \left|r_x^i - r_y^{\pi(i)}\right|\frac{c^p}{2}\right] + \frac{c^p}{2}\sum_{i = n_X + 1}^{n_Y}r_y^{\pi(i)} \nonumber \\ 
    &~~~ \left.+\left. \frac{c^p}{2}\sum_{i:d\left(p_x^i,p_y^{\pi(i)}\right)\geq c} \left(r_x^i + r_y^{\pi(i)}\right)\right)\right]^{\frac{1}{p}}.
\end{align}

Let $\gamma_\pi$ be an assignment set between the sequences $(1,\dots,n_X)$ and $(\pi(1),\dots,\pi(n_X))$, which satisfies $\gamma_\pi = \{(i,j):j = \pi(i) \text{~and~} d(p_x^i,p_y^j) < c\}$. Then we have that 
\begin{align}
    &d_p^{(c,2)}(f_X,f_Y) \nonumber \\ 
    &= \left[\min_{\pi \in \Pi_{n_Y}}\left(\sum_{(i,j)\in\gamma_\pi}\left[\min\left(r_x^i,r_y^{j}\right)d\left(p^i_x,p_y^{j}\right)^p \right.\right.\right. \nonumber \\
    &~~~\left. + \left|r_x^i - r_y^{j}\right|\frac{c^p}{2}\right] \nonumber\\
    &~~~\left. + \left.\frac{c^p}{2}\left(\sum_{i:\forall j,(i,j)\notin\gamma_\pi}r_x^i + \sum_{j:\forall i,(i,j)\notin\gamma_\pi}r_y^j\right)\right)\right]^{\frac{1}{p}}.
\end{align}
If $n_X > n_Y$, $d_p^{(c,2)}(f_X,f_Y) = d_p^{(c,2)}(f_Y,f_X)$. The general expression, independent of the sizes of $n_X$ and $n_Y$, is 
\begin{align}
    &d_p^{(c,2)}(f_X,f_Y) \nonumber \\ 
    &= \left[\min_{\pi \in \Pi_{\max(n_X,n_Y)}}\left(\sum_{(i,j)\in\gamma_\pi}\left[\min\left(r_x^i,r_y^{j}\right)d\left(p^i_x,p_y^{j}\right)^p \right.\right.\right. \nonumber \\
    &~~~\left. + \left|r_x^i - r_y^{j}\right|\frac{c^p}{2}\right] \nonumber\\
    &~~~\left. + \left.\frac{c^p}{2}\left(\sum_{i:\forall j,(i,j)\notin\gamma_\pi}r_x^i + \sum_{j:\forall i,(i,j)\notin\gamma_\pi}r_y^j\right)\right)\right]^{\frac{1}{p}}, \label{eq_permutation_pgospa}
\end{align}
where $\gamma_\pi$ is now an assignment set between the sequences $(1,\dots,\min(n_X,n_Y))$ and $(\pi(1),\dots,\pi(\min(n_X,n_Y)))$.

We note that when we consider all possible permutations $\pi$, the union of all $\gamma_\pi$ covers every possible pair $(i,j)$ that can be formed between the two sets $\{1,\dots,n_X\}$ and $\{1,\dots,n_Y\}$. This is because for any specific pair $(i,j)$, we can construct a permutation such that $\pi(i) = j$. This allows us to rewrite \eqref{eq_permutation_pgospa} as \eqref{eq_assignment_pgospa}.

This finishes the proof of Proposition \ref{prop_pgospa_alpha2}.

\bibliographystyle{IEEEtran}
\bibliography{references}

\vfill

\end{document}